\documentclass[a4paper,11pt]{article}
\pdfoutput=1 

\usepackage{jcappub} 
                     \usepackage[normalem]{ulem}
\usepackage[T1]{fontenc} 
\usepackage{aas_macros}
\usepackage[table]{xcolor}
\newcommand\y{\cellcolor{gray!20}}

\title{\boldmath Galaxy redshift-space bispectrum: the Importance of Being Anisotropic}

\author[a,b]{D. Gualdi}
\author[a,c]{L. Verde}

\affiliation[a]{Institut de Ciencies del Cosmos, University of Barcelona, ICCUB, Barcelona 08028, Spain}
\affiliation[b]{Institute of Space Studies of Catalonia (IEEC), E-08034 Barcelona, Spain}
\affiliation[c]{Instituci\'o Catalana de Recerca i Estudis Avan\c{c}ats, Passeig Lluis Companys 23, Barcelona 08010, Spain}

\emailAdd{dgualdi@icc.ub.edu}
\emailAdd{liciaverde@icc.ub.edu}

\abstract{
We forecast the benefits induced by adding the bispectrum anisotropic signal to the standard,  two- and three-point, clustering statistics analysis.
In particular, we forecast cosmological parameter constraints including the bispectrum higher multipoles terms together with the galaxy power spectrum (monopole plus quadrupole) and isotropic bispectrum (monopole) data vectors.
To do so, an analytical covariance matrix model is presented. This template is carefully calibrated on well-known terms of a numerical covariance matrix estimated from a set of simulated galaxy catalogues.
After testing the calibration using the power spectrum and isotropic bispectrum measurements from the same set of simulations, we extend the covariance modelling to the galaxy bispectrum higher multipoles.
Using this covariance matrix we proceed to perform cosmological parameter inference using a suitably generated mock data vector.
Including the bispectrum mutipoles up to the  hexadecapole, yields 1-D $68\%$ credible regions for the set of parameters $(b_1,b_2,f,\sigma_8,f_\mathrm{NL},\alpha_\perp, \alpha_\parallel)$ tighter by a factor of 30$\%$ on average for $k_\mathrm{max}=0.09\,h$/Mpc, significantly reducing at the same time the degeneracies present in the posterior distribution.}

\begin{document}
\maketitle
\flushbottom

\section{Introduction}
\label{sec:intro}

Higher-order statistics represent the most direct tool to access the cosmological information encrypted in the non-Gaussian features of the matter density field. Three-point (3pt) statistics are the lowest order, and therefore the  simplest, tools enabling us to harvest this information. The most recent analyses for the Baryon Oscillation Spectroscopic Survey (BOSS) \cite{Alam:2015mbd} spectroscopic data-set, part of the Sloan Digital Sky Survey (SDSS) \cite{Eisenstein:2011sa}, have been performed for both the 3pt correlation function \cite{Slepian:2015hca} and its Fourier transform, the bispectrum \cite{Gil-Marin:2016wya}.

Approximately four decades have passed since the first 3pt statistics studies \cite{Groth:1977gj,1975ApJ...196....1P,1982ApJ...259..474F,Fry:1983cj}. However only twenty years ago a mature formalism was developed to map theoretical models to spectroscopic data-sets in redshift space \cite{Matarrese:1997sk,Verde:1998zr,Verde:2001sf,Scoccimarro:1997st,Scoccimarro:2000ee,Scoccimarro:1999ed,Scoccimarro:2000sp}.

The window on non-linear scales was opened with the loop corrections inclusion \citep{Sefusatti:2009qh,Sefusatti:2011gt,Hashimoto:2017klo,Desjacques:2018pfv,Eggemeier:2018qae,Castiblanco:2018qsd}; higher-order terms become important at non-linear scales and are indeed necessary to study primordial non-Gaussianity using the 3pt statistics to lift the degeneracy with the other cosmological parameters \citep{Verde:1999ij,Scoccimarro:2003wn,Jeong:2009vd,Bose:2018zpk,Karagiannis:2018jdt}.

The general benefits of using the bispectrum in addition to the power spectrum have recently been studied in \cite{Yankelevich:2018uaz,Oddo:2019run,Barreira:2019icq}. Originally, the bispectrum was thought to be useful in order to break the degenaracies between bias and cosmological parameters \cite{Fry:1992vr,Fry:1993bj}. Lately, the bispectrum has been shown to be a key tool for reaching more specific goals, such as detecting General Relativity effects \cite{GilMarin:2011xq,Bartolo:2013ws,Bellini:2015wfa,Bertacca:2017dzm,DiDio:2018unb} and improving the neutrino masses  constraints \cite{Ruggeri:2017dda,Coulton:2018ebd,Hahn:2019zob}.

Nevertheless, all the analyses performed to date on spectroscopic data have been limited to the isotropic signal component, the so-called bispectrum "monopole" \cite{Gil-Marin:2014sta,Gil-Marin:2016wya}. In analogy to the power spectrum, redshift space distortions (RSD) induce an  anisotropic signal with respect to the line of sight (see pioneering work by  \cite{Kaiser:1987qv,Hamilton:1997zq}). This anisotropy can be expanded  in multipole moments, where the monopole is just the averaged signal in all directions around the line of sight. To date, multipoles beyond the monopole have  been used to constrain cosmological parameters only for the power spectrum \cite{Montesano:2010qc,Montesano:2011bp,Beutler:2013yhm,Blake:2018tou}, with the state-of-the-art at the moment of writing being Refs. \cite{Beutler:2016arn,Zheng:2018kgq,Ivanov:2019pdj,Ivanov:2019hqk,Philcox:2020vvt,Colas:2019ret}. Measuring higher-order  multipoles of the redshift-space galaxy bispectrum is complicated in terms of estimators modelling \cite{Scoccimarro:2015bla,Hashimoto:2017klo,Sugiyama:2018yzo}, even when simulated volumes with periodic boundary conditions are used. For galaxy surveys, these measurements are even more difficult because of the necessity of accounting for all observational effects (e.g., window function effect, varying average density number, varying line of sight, etc.). 
As in the case of the power spectrum, the anisotropic signal in the bispectrum due to RSD potentially carries key additional cosmological information.

This additional signal can be used not only to obtain tighter parameter constraints, but also as a self-consistency check for the theoretical model itself. In other words one could test whether the same set of cosmological parameters can produce an analytical template that fits well all the measured data vector components (power spectrum / bispectrum, isotropic / anisotropic).
Hence we have a strong motivation to investigate the potential of the   galaxy bispectrum multipoles.

The purpose of this work is to assess the added constraining power obtained by including higher-order multipoles of the bispectrum to the so far used large scale structure (LSS) statistics in Fourier space (power specrum multipoles and bispectrum monopole).
We study the improvements, in terms of tighter parameter constraints, in the 1- and 2-D $68\%$ credible regions resulting from using the bispectrum multipoles. Fisher forecasts for the bispectrum multipoles, using a diagonal covariance matrix including only Gaussian terms, were realised in \cite{Gagrani:2016rfy}. Here we go beyond this approximation, dedicating a significant effort in modelling the off-diagonal, non-Gaussian terms of the covariance matrix.

Since measurements of these new statistics (the anisotropic bispectrum modulation) either from mocks or data are at the moment extremely expensive and the necessary codes are not fully developed or publicly available yet, it is important to estimate first the potential gain these new statistics could offer. To do so however the full covariance matrix must be estimated in order to derive the cosmological parameters posterior distributions, for example through Markov Chain Monte Carlo methods (MCMC).  
We argue here that resorting to MCMC's  is indeed necessary to  explore and place constraints on parameters-space  where there are non-trivial degeneracies, such as the parameters usually constrained in a clustering analysis. Fisher forecast can not do this since are based on the first derivatives of the data-vector's theoretical model.

Note that the covariance must be estimated for the full "data vector" which comprises both the power spectrum and bispectrum multipoles;  this matrix cannot be approximated as diagonal. Traditionally, there are two approaches to estimate covariance matrices: from a large number of mocks or analytically. Adopting the first approach alone is not viable here, for the reason explained above that measurements of these statistics from a large enough set of mocks are not currently available. 
Therefore we  model the covariance matrix analytically, calibrating it, as much as possible, for the terms that are measurable, on estimates from a large number of mocks. We then use this analytical template, calibrated  and extended to terms involving the bispectrum higher multipoles, to evaluate the likelihood in the MCMC sampling of the parameters posterior distribution.

An  analytical model of the galaxy power spectrum and bispectrum multipoles covariance matrix  alternative to the one presented here, was recently derived in \cite{Sugiyama:2019ike}, where the effect of including the anisotropic bispectrum signal was studied through the cumulative signal to noise ratio.

In order to model the galaxy bispectrum higher multipoles we use the choice of angles described in \cite{Hashimoto:2017klo} which differs from the one recently proposed in \cite{Sugiyama:2018yzo}. The expansion is then performed in terms of the angle between the normal to the triangle's surface and the line of sight, together with the angle describing the rotation of the triangle on the plane defined by its sides (or in other words the rotation around the normal to the triangle's surface).

The structure of the paper is the following.
In Section \ref{sec:problem_objectives} we give an overview of the objectives and the necessary steps to achieve them.
In Section \ref{sec:dv_model} we describe the analytical expressions used to model the data vector while in Section \ref{sec:cov_matrix_model} we list the analytical terms that we included in our covariance matrix model.
Section \ref{sec:compare_covs} is of critical importance since it describes the qualitative and quantitative tests that we performed in order to check the accuracy and reliability of our covariance analytical model. In Section \ref{sec:higher_bk_multi} we forecast through MCMC sampling the parameter constraints improvements obtained by adding to the data vector the galaxy bispectrum higher multipoles. We conclude in Section \ref{sec:conclusions}.

\begin{figure}[tbp]
\centering 
\includegraphics[width=.90\textwidth]
{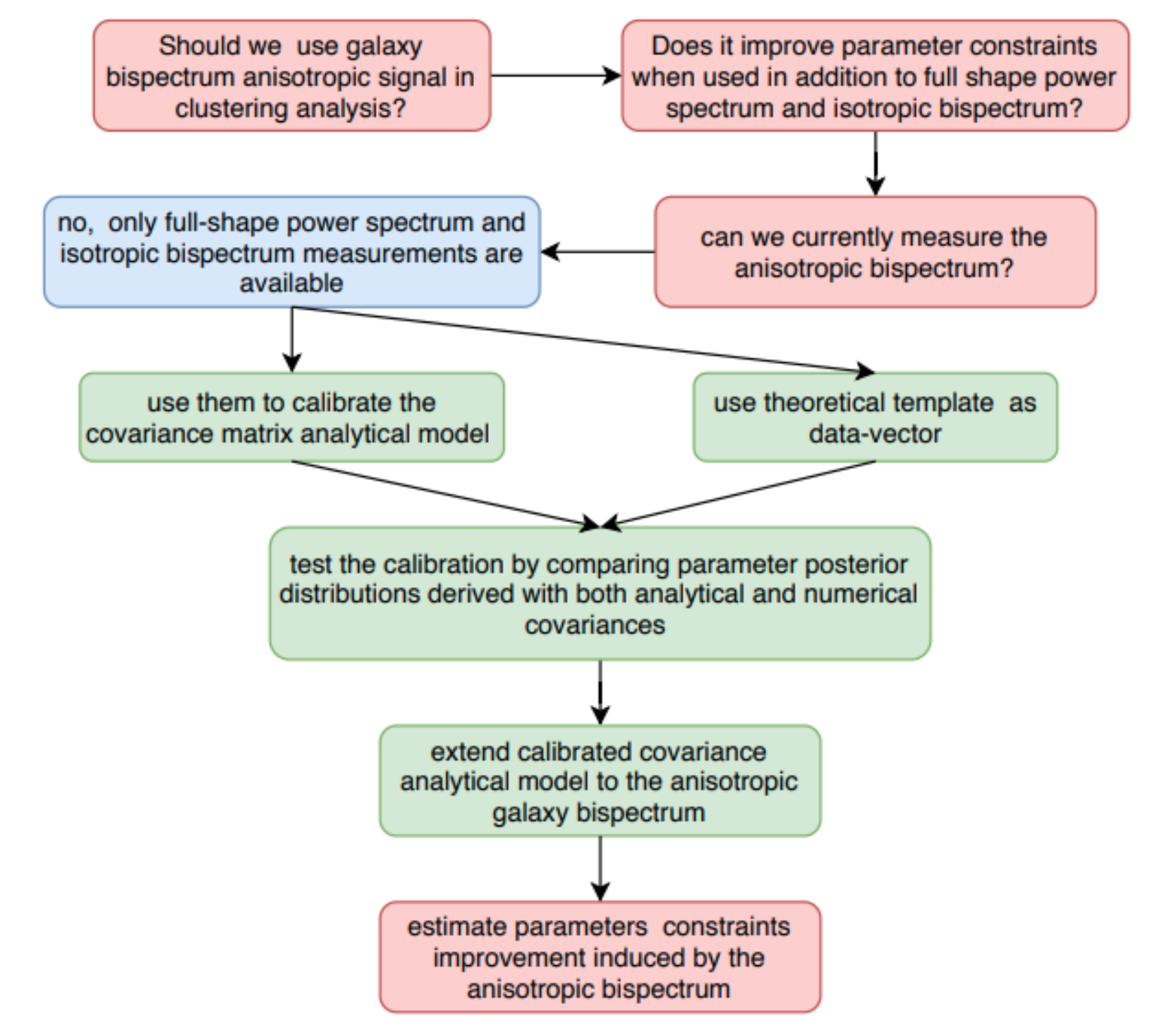}
\caption{\label{fig:conceptual_map}Conceptual map of the paper.}
\end{figure}

\section{Statement of the problem, objectives and roadmap}
\label{sec:problem_objectives}
It is not at all a trivial task to develop the estimators and the associated numerical pipelines to measure the galaxy-redshift space bispectrum anisotropic signal in terms of multipoles expansion  \cite{Scoccimarro:2015bla,Sugiyama:2019ike}.
Hence it is important to ask: is it worth it?
There are two aspects to this question. One is the simple potential in reducing parameters errors or lifting parameters degeneracies. The other aspect is that in any practical application,  these extra statistics can offer a powerful consistency check of the modelling and thus  possibly of the model itself. The second aspect, while possibly of more consequence than the first, cannot be easily demonstrated and quantified before one is  able to measure these statistics on mock survey catalogs ({\it mocks} for short). However,   it is reasonable to expect that these statistics can provide a useful consistency checks if they carry significant additional information, and therefore only if they are useful to reduce parameters errors.    
This is what we set up to address in  this work.

Our goal is indeed to show whether adding the bispectrum anisotropic signal to the standard power spectrum $\mathrm{P}_\mathrm{g}^{(\ell)}$ (isotropic + anisotropic) and bispectrum $\mathrm{B}_\mathrm{g}^{(00)}$ (isotropic) data vector, produces tighter 1-2D marginalised posterior distributions for the parameters usually constrained by galaxy-clustering analyses. We prove this through the path described  in the conceptual map of Figure~ \ref{fig:conceptual_map}.

Given the absence of measurements, the first task consists in analytically modelling the bispectrum anistropic signal.  We choose a decomposition in terms of Legendre polynomials for a specific choice of angles describing the orientation with respect to the line of sight. We  then proceed to use this theoretical model as a "mock"\footnote{The meaning of "mock" here is different from the more widespread use of the term. Usually "mock" data refer to  carefully simulated data in such a way as to have properties as similar as possible to the real data. Here our generated data vector is somewhat idealized, being obtained from theoretical templates. For this reason we use the term "mock" here in quotes. In the rest of the paper we will  use the word mock in its more common meaning and refer for brevity to the template-generated  data as (fiducial) template data vector. In some applications we will use a data vector obtained from the survey mocks. We will refer to this as simulated data vector.} data vector. A similar approach was used in \cite{Perotto:2006rj,Brinckmann:2018owf} to derive forecasts on the sum of the neutrino masses constraints.

In summary our "data" vector is composed by four blocks, galaxy power spectrum (isotropic plus anisotropic) and galaxy bispectrum (also isotropic plus anisotropic):\\ $\left[\mathrm{P}^{(0)}_\mathrm{g};\mathrm{P}^{(2)}_\mathrm{g};\mathrm{B}^{(00)}_\mathrm{g};\mathrm{B}^{(\imath\jmath)}_\mathrm{g}\right]$, where $\mathrm{B}^{(\imath\jmath)}_\mathrm{g}$ indicates all the possible bispectrum higher multipoles. Each power spectrum block includes all the $k$-bins (bandpowers) while each bispectrum block contains all the triangle configurations used.

As it is widely done in large-scale structure studies, we assume the likelihood of our data vector to be Gaussian, with a covariance matrix that does not depend on the cosmological parameters. This assumption has been extensively discussed, tested, and motivated, see e.g., \cite{Carron:2012pw,Repp:2015jja} for an insightful discussion. 

To perform likelihood-based parameter inference, 
 we also need the covariance matrix.  The covariance matrix  is usually estimated numerically from the data vector's measurements on a large enough set of simulations or mock survey data. 
In our case we only have measurements for a reduced data vector which includes the standard galaxy clustering statistics, $\left[\mathrm{P}_\mathrm{g}^{(\ell)};\mathrm{B}_\mathrm{g}^{(00)}\right]$ and therefore only for a sub-matrix.  To overcome this obstacle, we derive an analytical model for the full covariance matrix to be used as a template. This template can then be calibrated using the numerically estimated covariance for the sub matrix involving $\left[\mathrm{P}_\mathrm{g}^{(\ell)};\mathrm{B}_\mathrm{g}^{(00)}\right]$, for which we have measurements from 1400 galaxy mock catalogues. The calibration is done by fitting parameters which are common to both the anisotropic and isotropic components of the  full data vector's covariance and therefore independent of the multipole expansion.
Two of them are amplitudes parameters (one for each of the power spectrum and bispectrum parts respectively) and the third one is an effective average galaxy number density (modelling the shot-noise). The fit is done by comparing our analytical template to the covariance matrix diagonal estimated from the mocks. For the calibration we only use the diagonal elements since  they are signal-dominated (i.e.,  their statistical noise is negligible with respect to the physical signal).
A similar approach was followed by Ref.~ \cite{Slepian:2015hca} to compute the three-point correlation function's covariance matrix .
In the analytic expression for the covariance matrix it is hard to include the  extra mode-coupling and  coupling between different statistics, induced by the window. This will be taken into consideration when performing the  comparison to the numerical estimate of the covariance.

We use the analytical model for the  power spectrum multipoles covariance only for the calibration process.  In our forecasts,  in order to minimize potential systematic errors connected with the power spectrum covariance modelling, for this part of the data vector, $\left[\mathrm{P}_\mathrm{g}^{(\ell)}\right]$, we will use the covariance numerically estimated from the simulations. The resulting "hybrid" covariance model will be referenced as HYB in the figures.

The following step is to verify that our covariance analytical model, when used to sample via a Markov chain Monte Carlo (MCMC)  the parameters posterior distribution, produces the same results obtained with the  numerical covariance matrix. 

We test this for the case of the anisotropic power spectrum plus isotropic bispectrum data vector (for which we have measurements from a large set of galaxy mock catalogues).
In particular we compare the posterior distributions 1D medians and $68\%$ credible intervals  for a set of 1400 galaxy mock catalogues.

Provided that there is no significant difference between the parameters posteriors obtained using  the analytical covariance matrix template and the numerically estimated one, we  can  assume that the calibrated covariance model can  safely be extended  to the bispectrum multipoles.

The analytical model of the extended covariance matrix can  then be used to evaluate the effect  on cosmological parameter inference of including  higher  bispectrum multipoles in  the data vector.

We will show in Section \ref{sec:improvement_parameters} two main  results. First we prove that the anisotropic bispectrum signal contains additional significant cosmological information. Second,  this improvement saturates at the level of the hexadecapole terms, meaning that measuring higher multipoles  is not expected to  further  tighten parameter constraints (i.e., it would not be worth the effort).

\section{Signal modelling}
\label{sec:dv_model}
Given the lack of measurements from simulations or "real world" surveys, in our work we simulate the bispectrum anisotropic signal using an analytical template. Therefore we employ theoretical expressions derived from standard perturbation theory (see \cite{Bernardeau:2001qr} for a complete review) to generate our "fiducial" (but realistic) template data vector.

 The tree-level expression for the power spectrum and bispectrum are:
\begin{align}
\mathrm{P}_{\mathrm{g}}\left(\mathbf{k}\right)&=\mathrm{Z^\mathrm{s}_1}\left[\mathbf{k}\right]^2\mathrm{P}_{\mathrm{m}}\left( k \right)D_{FoG}^P(k)
\notag\\
\mathrm{B}_{\mathrm{g}}\left(\mathbf{k}_1,\mathbf{k}_2,\mathbf{k}_3\right)\,&=\,\left[2\,\mathrm{P}_{\mathrm{m}}\left(k_1\right)\mathrm{P}_{\mathrm{m}}\left(k_2\right)\mathrm{Z^\mathrm{s}_1}\left[\mathbf{k}_1\right]\mathrm{Z^\mathrm{s}_1}\left[\mathbf{k}_2\right]\mathrm{Z^\mathrm{s}_2}\left[\mathbf{k}_1,\mathbf{k}_2\right] \right.
\notag \\
&+\,\left. \mathrm{two\,cyclic\, terms}\right]\times  D_{FoG}^B(k_1,k_2,k_3),
\end{align}

\noindent where the subscript "g" indicates that the statistics are computed for the galaxy field in redshift space. The standard perturbation theory kernels in redshift space used in this work ($\mathrm{Z}^\mathrm{s}_n$) are those reported in Appendix C of \cite{Gil-Marin:2014sta}.
The $D_{FoG}$ terms model the Fingers-of-God effect (hereafter FoG) \cite{Jackson:2008yv} introducing two effective parameters $\sigma_\mathrm{P}$ and $\sigma_\mathrm{B}$ for the power spectrum and bispectrum respectively. These parameters enter into two Lorentzian damping functions in front of the respective data vectors \cite{Gil-Marin:2014sta}:

\begin{align}
\label{eq:fog_parameters}
   D_{FoG}^P&=\dfrac{1}{1 + k^2\mu^2\sigma_\mathrm{P}^2/2 }\, \notag \\
    D_{FoG}^B&=\,\dfrac{1}{1 + (k_1^2\mu_1^2+k_2^2\mu_2^2+k_3^2\mu_3^2)\sigma_\mathrm{B}^2/2 }\, .
\end{align}{}

\noindent Beside depending on the $k$-vectors coordinates, the $\mathrm{Z_n^s}$ kernels are also a function of the bias parameters $b_1$, $b_2$ and $b_{s^2}$ together with the growth rate $f$. 

As is it  widespread, we also make the approximation of the bias being local in Lagrangian space, which allows us to express the non-local bias term $b_{s^2}$ as function of the linear bias parameter $b_1$, $b_{s^2}=-\dfrac{4}{7}(b_1-1)$ \cite{Chan:2012jj,Baldauf:2012hs,Gil-Marin:2014sta}.

In our modeling and analysis we include the Alcock-Paczy\`{n}ski \cite{Alcock:1979mp} parameters to account for differences between the true underlying cosmology and the fiducial one used to convert redshift measurements into Euclidean distances. A discrepancy between the fiducial and true underlying cosmology induces distortions in the clustering properties. This can be modelled through the parallel and perpendicular dilation parameters  $\alpha_\parallel = k_\parallel/p_\parallel$ and $\alpha_\perp = k_\perp/p_\perp$ which relate the measured wave-vector $\mathbf{k}$ with the corresponding wave-vector $\mathbf{p}$ for a particular cosmology. Being $\mu$ and $\eta$ the cosines of the angles between the line of sight and $\mathbf{k}$ and $\mathbf{p}$ respectively, the conversion is given by \cite{Gil-Marin:2016wya}

\begin{align}
    \label{eq:ap_par_rels}
    p &= |\mathbf{p}| = \dfrac{k}{\alpha_\perp}\left[1+\mu^2\left(F^{-2}-1\right)\right]^{\frac{1}{2}} 
    \notag \\
    \eta &= \dfrac{\mu}{F}\left[1+\mu^2\left(F^{-2}-1\right)\right]^{-\frac{1}{2}}\,,
\end{align}

\noindent where $F\equiv\alpha_\parallel / \alpha_\perp$.

\subsection{Multipole expansion}
We compute the multipole expansion of the tree-level galaxy power spectrum  with respect to the line of sight in the standard way

\begin{eqnarray}
\label{eq:pk_model}
\mathrm{P}_\mathrm{g}^{(\ell)}(k) = \dfrac{(2\ell + 1)}{2\alpha_\parallel\alpha_\perp^2}\int^{+1}_{-1}d\mu \,\mathcal{L}_\ell(\mu)\,\mathrm{P}_\mathrm{g}(p,\eta)\,
\end{eqnarray}{}

\noindent where $\mu$ is the cosine of the angle between the line of sight and the $k$-vector while $\mathcal{L}_\ell(\mu)$ is the $\ell$-th order Legendre polynomial.

For the bispectrum multipole expansion we adopt the choice of angles presented in \cite{Hashimoto:2017klo}. The first angle is the one between the normal to the surface of the triangle and the line of sight, $\hat{\mathbf{z}}$, whose cosine we indicate by the letter $\mu$. The second angle describes the rotation of the triangle around the normal to its surface, with a fantasy effort we call $\nu$ its cosine;

\begin{eqnarray}
\mu = \cos{\omega} &=& \dfrac{\left(\mathbf{k}_1\times \mathbf{k}_2\right)\cdot\hat{\mathbf{z}}}{\sin{\theta_{12}}}
\notag \\
\nu = \cos{\phi} &=& \dfrac{\left[\hat{\mathbf{z}}\times\left(\mathbf{k}_1\times \mathbf{k}_2\right)\right]\cdot \hat{\mathbf{k}}_1}{\sin{\omega}}\,,
\end{eqnarray}{}

\noindent where $\cos{\theta_{12}}=(\mathbf{k_1}\cdot\mathbf{k}_2)/(k_1k_2)$. 
The origin of both angles can be for example imagined translating the line of sight and normal to the surface vectors to $\mathbf{k}_1$'s starting point (see Figure 1 of \cite{Hashimoto:2017klo} for a visual representation).

For numerical reasons we still integrate over $\mu_1$ and $\mu_2$ (the cosine of the angles between the line of sight and the $k$-vectors $k_1$ and $k_2$) since the $\mu$ and $\nu$ are derived quantities from the standard bispectrum coordinates $(k_1,k_2,k_3,\mu_1,\mu_2)$, conversion formula given in  Appendix \ref{sec:appendixB2}, Equation \ref{eq:angles_conversion}. The bispectrum multipoles are then given by

\begin{eqnarray}
\label{eq:bk_model}
\mathrm{B}_\mathrm{g}^{(\imath,\jmath)}(k_1,k_2,k_3) = \dfrac{(2\imath+1)(2\jmath+1)}{4\pi \alpha_\parallel^2\alpha_\perp^4}\int^{+1}_{-1}d\mu_1\int^{2\pi}_0 d\phi \,
\mathcal{L}_\imath(\mu)\mathcal{L}_\jmath(\nu)\mathrm{B}_\mathrm{g}(p_1,p_2,p_3,\eta_1,\eta_2)\,,
\end{eqnarray}{}

\noindent where $\mathcal{L}_\imath(\mu)$ and $\mathcal{L}_\jmath(\nu)$ are the $\imath$-th and $\jmath$-th order Legendre polynomials for $\mu$ and $\nu$, respectively. The azimuthal angle $\phi$ around $\mathbf{k}_1$ is defined in Section 3 Equation 18 of \cite{Scoccimarro:1999ed} as $\mu_2\equiv\mu_1\cos\theta_{12} - \sqrt{1-\mu_1^2}\sqrt{1-\cos\theta_{12}^2}\cos\phi$, which was used also in \cite{Gil-Marin:2014sta}.  $p_1$, $p_2$, $p_3$ are the $k$-modules while $\eta_1$, $\eta_2$ are the cosines of the angles between $\mathbf{k}_1$, $\mathbf{k}_2$ and the line of sight corrected for the Alcock-Paczy\`{n}ski effect (Equation \ref{eq:ap_par_rels}).

\section{Covariance matrices: formulae}
\label{sec:cov_matrix_model}
We update and extend the analytical results presented in \cite{Gualdi:2017iey,Gualdi:2018pyw}. The full derivation  and full expressions of the terms used in this section  are shown in Appendix \ref{sec:app_cov_terms}.
 The covariance matrix can be divided into different blocks which, given the definition of the data vector components in Equations \ref{eq:pk_model} and \ref{eq:bk_model} we organize as follows:

\begin{eqnarray}
\label{eq:cov_matrix}
\begin{pmatrix}
\y \mathrm{C}^{\mathrm{P}^{0}_{\mathrm{g}}} & \y \mathrm{C}^{\mathrm{P}^{0}_{\mathrm{g}}\mathrm{P}^{2}_{\mathrm{g}}} & \y \mathrm{C}^{\mathrm{P}^{0}_{\mathrm{g}}\mathrm{B}^{00}_{\mathrm{g}}} & 
\mathrm{C}^{\mathrm{P}^{0}_{\mathrm{g}}\mathrm{B}^{\imath\jmath}_{\mathrm{g}}} \\
\y \mathrm{C}^{\mathrm{P}^{2}_{\mathrm{g}}\mathrm{P}^{0}_{\mathrm{g}}} & \y \mathrm{C}^{\mathrm{P}^{2}_{\mathrm{g}}} &  \y \mathrm{C}^{\mathrm{P}^{2}_{\mathrm{g}}\mathrm{B}^{00}_{\mathrm{g}}} & 
\mathrm{C}^{\mathrm{P}^{2}_{\mathrm{g}}\mathrm{B}^{\imath\jmath}_{\mathrm{g}}} \\
\y \mathrm{C}^{\mathrm{P}^{0}_{\mathrm{g}}\mathrm{B}^{00}_{\mathrm{g}}} & \y \mathrm{C}^{\mathrm{P}^{2}_{\mathrm{g}}\mathrm{B}^{00}_{\mathrm{g}}} &  \y \mathrm{C}^{\mathrm{B}^{00}_{\mathrm{g}}} & 
\mathrm{C}^{\mathrm{B}^{00}_{\mathrm{g}}\mathrm{B}^{\imath\jmath}_{\mathrm{g}}}  \\
\mathrm{C}^{\mathrm{P}^{0}_{\mathrm{g}}\mathrm{B}^{\imath\jmath}_{\mathrm{g}}} & \mathrm{C}^{\mathrm{P}^{2}_{\mathrm{g}}\mathrm{B}^{\imath\jmath}_{\mathrm{g}}} &  
\mathrm{C}^{\mathrm{B}^{\imath\jmath}_{\mathrm{g}}\mathrm{B}^{00}_{\mathrm{g}}} & 
\mathrm{C}^{\mathrm{B}^{\imath\jmath}_{\mathrm{g}}}  \\
\end{pmatrix}
\end{eqnarray}{}

\noindent In the above covariance matrix sketch on the diagonal there are  blocks for both the power spectrum $\mathrm{C}^{\mathrm{P}^{0}_{\mathrm{g}}}$ and bispectrum multipoles $\mathrm{C}^{\mathrm{B}^{\imath\jmath}_{\mathrm{g}}}$ auto-covariances. Off-diagonal blocks correspond to  the cross-covariances, between different multipoles of the same statistic, $\mathrm{C}^{\mathrm{P}^{0}_{\mathrm{g}}\mathrm{P}^{2}_{\mathrm{g}}}$ or $\mathrm{C}^{\mathrm{B}^{\imath\jmath}_{\mathrm{g}}\mathrm{B}^{00}_{\mathrm{g}}}$, and between different statistics, like $\mathrm{C}^{\mathrm{P}^{0}_{\mathrm{g}}\mathrm{B}^{\imath\jmath}_{\mathrm{g}}}$.
 The gray background indicates the covariance part that can be evaluated both analytically and numerically from the mock catalogues and therefore it can be used to calibrate the analytic expressions.

In all the auto-covariances and cross-covariances between different multipoles of the same statistic there are both Gaussian and non-Gaussian terms. The Gaussian terms are the ones obtained at lower order in perturbation theory, when the field is considered to be completely Gaussian and therefore all moments higher than the variance vanish. The Gaussian terms are non-zero only on the diagonal of each block. Non-Gaussian terms appear when higher-orders of the matter-density field are considered not only in modelling the data vector (where they are dominant for example in the case of the bispectrum), but also in the covariance matrix theoretical modelling. Without second-order terms (inducing a non-zero bispectrum) for example, there would not be any correlation between power spectrum and bispectrum data vector elements.

\begin{figure}[tbp]
\centering 
\includegraphics[width=.525\textwidth]
{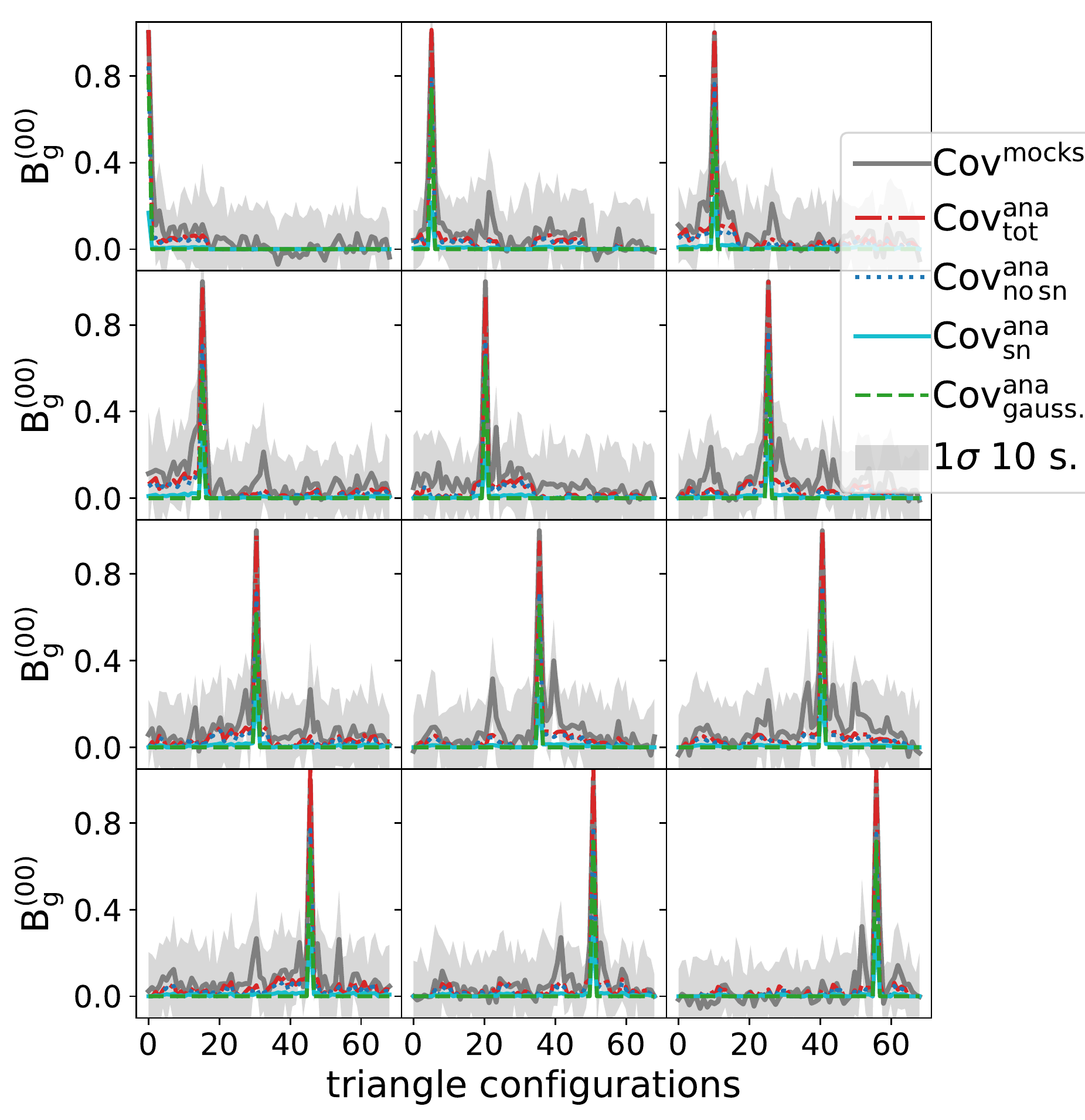}
\hfill
\includegraphics[width=.465\textwidth]
{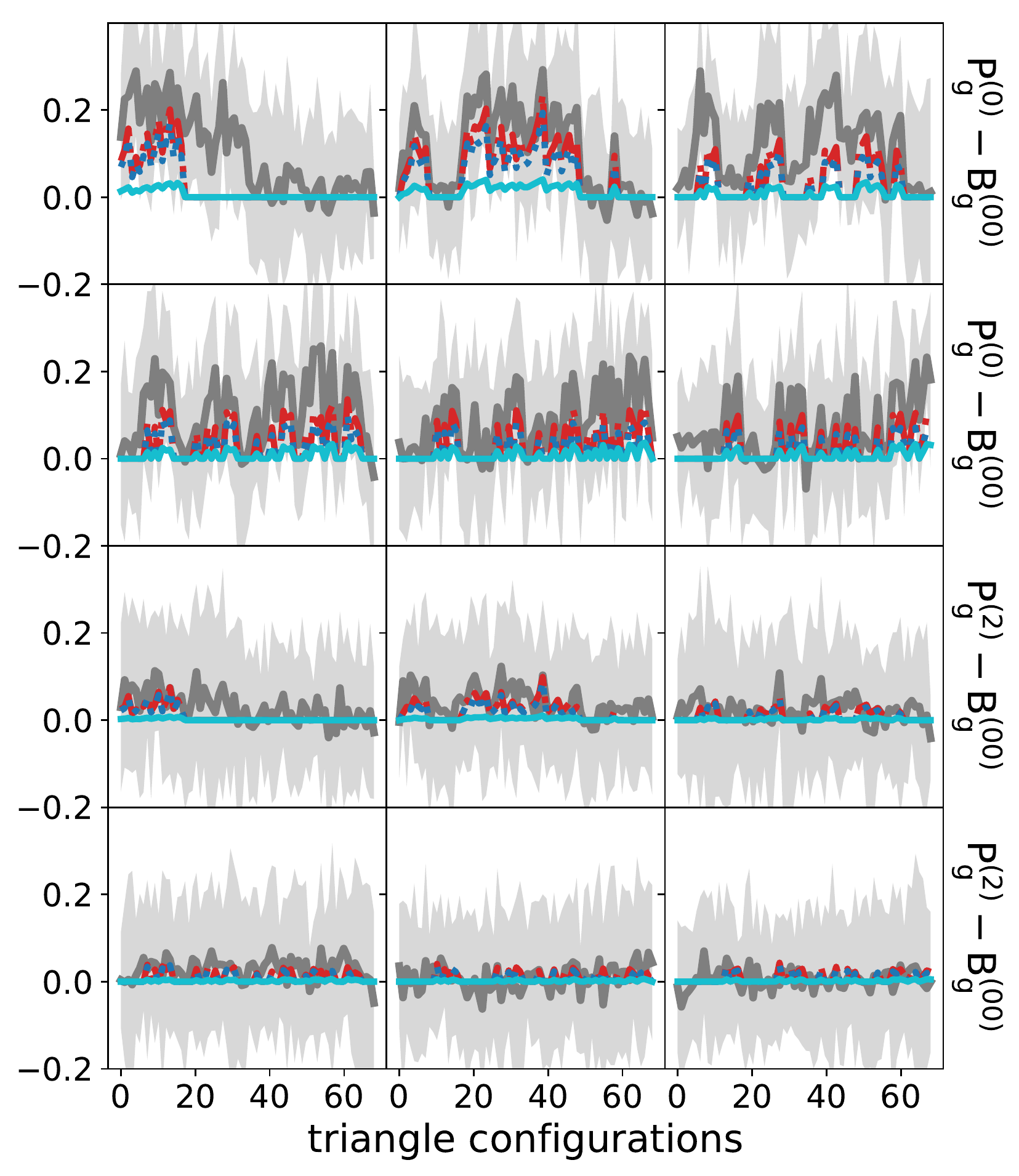}
\caption{\label{fig:covariance_comparison1D} Columns of the reduced covariance matrices, both analytical and numerical. The shaded area is obtained by randomly splitting the 1400 mocks in 10 groups, obtaining the numerical covariance matrix for each of them (correcting by the appropriate Hartlap factor) and finally computing each covariance matrix element standard deviation.
In the left panel is shown the comparison for the bispectrum auto-covariance matrix while on the right for its cross-correlation with each of the power spectrum multipoles modes. 
The different lines shows the contributions given by the Gaussian term, the shot-noise correction and the non-Gaussian contribution.}
\end{figure}

Differently from Ref. \cite{Sugiyama:2019ike}, we limit our perturbation theory expansion of the non-Gaussian terms to the lowest order for each component of the covariance matrix ( $\propto\delta_\mathrm{m}^6$ in the cross-covariance and $\propto\delta_\mathrm{m}^8$ in the bispectrum auto-covariance, $\delta_\mathrm{m}$ being the matter perturbation density variable). This implies not modelling higher-order terms such as the one proportional to the tetraspectrum ($\propto\delta_\mathrm{m}^8$) in the cross-covariance between power spectrum and bispectrum together with the one proportional to the pentaspectrum ($\propto\delta_\mathrm{m}^{10}$) in the bispectrum auto covariance.  According to  Figures 2 and 3 of Ref.~\cite{Sugiyama:2019ike} for our range of scales ($k<0.09$h Mpc$^{-1}$) and considering the scatter in the covariance matrix terms we will later discuss in Figure \ref{fig:covariance_comparison1D}, the lack of the above mentioned higher-order terms is expected to  have negligible impact in our analysis. Moreover, since we calibrate the amplitude of the covariance  terms on a suite of mocks (see Sec.~\ref{sec:compare_covs}), any possible residual is likely absorbed by this normalization.
Full expressions and more extended discussion can be found in Appendix \ref{sec:app_cov_terms}.

For the covariance between different galaxy power spectrum multipoles we  consider only the Gaussian term since on linear scales  it is dominant compared  to the non-Gaussian part  \cite{Scoccimarro:1999kp}. In particular, the covariance between the power spectrum $\ell_1$-multipole for mode $k_1$ and the power spectrum $\ell_2$-multipole for mode $k_2$ is given by

\begin{align}
\label{eq:cov_pp}
&\mathrm{C}_\mathrm{G}^{\mathrm{P}^{\ell_1}_{\mathrm{g}}\mathrm{P}^{\ell_2}_{\mathrm{g}}}\left(k_1;k_2\right) = 
\left(2\ell_1+1\right)\left(2\ell_2+1\right)
\dfrac{2(2\pi)^3\delta^\mathrm{K}_{12}}{4\pi k_1^2\Delta kV_\mathrm{s}}\quad
\dfrac{1}{2}\int^{1}_{-1}d\mu_1 \mathcal{L}_{\ell_1}\left(\mu_1\right)\mathcal{L}_{\ell_2}\left(\mu_1\right)
\mathrm{P}_{\mathrm{g}}\left(k_1,\mu_1\right)^2,
\end{align}

\noindent where $\ell_1,\ell_2=0,2$. $V_\mathrm{s}$ is the survey effective volume (where for the analytical model we approximate $V_\mathrm{s}=(2\pi)^3k_f^{-3}$), $\Delta k$ is chosen resolution in Fourier space for the power spectrum data vector. 

In order to correct for the shot-noise, in all our modelling we implicitly adopt the expressions derived in \cite{Sugiyama:2019ike}, for example $\mathrm{P}_{\mathrm{g}}\left(k_1,\mu_1\right)\equiv \mathrm{P}^N_{\mathrm{g}}\left(k_1,\mu_1\right)=\mathrm{P}_{\mathrm{g}}\left(k_1,\mu_1\right) + 1/n_\mathrm{g}$. This  is appropriate because the estimators used to measure the power spectrum and bispectrum statistics from galaxy catalogues and data have the Poisson shot-noise subtraction implemented. However the shot-noise contribution must still be included in the covariance matrix.

The Kronecker's delta  $\delta^\mathrm{K}_{12}$ is selecting only equal $k$'s (in the sense of $|k_a-k_b|<\Delta k/2$). The above basic model for the power spectrum multipoles covariance matrix diagonal only serves to the purpose of fitting the amplitude parameter needed for the cross-correlation term between power spectrum and bispectrum as described later in Section \ref{sec:compare_covs}. Indeed in our analytical covariance matrix, for the power spectrum multipoles auto-covariance, we use the one estimated from the mocks, hence the name "hybrid" covariance matrix (referenced as HYB in the figures). In this way we avoid any possible source of systematic error linked with a non-optimal modelling of the power spectrum covariance. For a recent specific work on the accurate modelling of the power spectrum multipoles covariance matrix see \cite{Wadekar:2019rdu}.

In order to deal with the many indexes needed for the galaxy bispectrum multipole expansion we introduce the following notation.
Let's say we want to  compute the covariance between two  different bispectrum configurations $a$, $b$, and  multipoles, $\mathrm{B}_\mathrm{g}^{(\alpha\beta)}$ and $\mathrm{B}_\mathrm{g}^{(\kappa\lambda)}$. The indices $\alpha$, $\beta$ etc. correspond to the multipoles $\ell=0,2,4$, where the first index refers to the angle cosine $\mu$ and the second to $\nu$. 
Therefore the Legendre polynomials for both angle cosines $\mu$ and $\nu$ will be written as $\mathcal{L}^{(\alpha)}_{\mu_a}$, $\mathcal{L}^{(\beta)}_{\nu_a}$ for the  triangle $a$, and $\mathcal{L}^{(\kappa)}_{\mu_b}$, $\mathcal{L}^{(\lambda)}_{\nu_b}$ for the triangle $b$. For the sake of brevity we introduce the short notation $\mathcal{M}^{(\alpha\beta\kappa\lambda)}_{\mu\nu,ab} =\mathcal{L}^{(\alpha)}_{\mu_a}\mathcal{L}^{(\beta)}_{\nu_a}\mathcal{L}^{(\kappa)}_{\mu_b}\mathcal{L}^{(\lambda)}_{\nu_b} $ with the added benefit of appearing cool general relativity-like. In the same way we write the compact pre-factor $\mathcal{C}^{\alpha\beta}_{\kappa\lambda} = (2\alpha+1)(2\beta+1)(2\kappa+1)(2\lambda+1)$.

The bispectrum covariance matrix at lowest order in terms of matter overdensity field perturbative expansion is composed of three terms: the first is proportional to the product of three power spectra (Gaussian), the second and the third are proportional to the product of two bispectra and power spectrum times trispectrum (non-Gaussian), respectively. The six-points correlator generating these contributions is reported in Equation \ref{eq:bk_cov_expansion}. 
The galaxy bispectrum multipoles covariance Gaussian term (i.e., first line of Eq. \ref{eq:bk_cov_expansion}) is then given by

\begin{align}
\label{eq:cov_bp_gg}
\mathrm{C}_{\propto\mathrm{P}^3}^{\mathrm{B}^{\alpha\beta}_{\mathrm{g}}\mathrm{B}^{\kappa\lambda}_{\mathrm{g}}}\left(k_1,k_2,k_3;k_4,k_5,k_6\right) &=
\dfrac{\mathrm{D}^{123}_{456} (2\pi)^6\mathcal{C}^{\alpha\beta}_{\kappa\lambda}}
{V^{\mathrm{t}}_{123}V_\mathrm{s}}
\notag \\
&\times\dfrac{1}{4\pi}
\int^{+1}_{-1}d\mu_1\int^{2\pi}_{0}d\phi
\mathcal{M}^{(\alpha\beta\kappa\lambda)}_{\mu\nu,aa}
\mathrm{P}_{\mathrm{g}}\left(k_1,\mu_1\right)\mathrm{P}_{\mathrm{g}}\left(k_2,\mu_2\right)\mathrm{P}_{\mathrm{g}}\left(k_3,\mu_3\right),
\end{align}{}

\noindent where $\mathrm{D}^{123}_{456}$ stands for all the possible indexes permutations of the product $\delta^\mathrm{K}_{14}\delta^\mathrm{K}_{25}\delta^\mathrm{K}_{36}$ and has values $6,2,1$ respectively for equilateral, isosceles and scalene triangles. Finally $V^{\mathrm{t}}_{123}=8\pi^2k_1k_2k_3\Delta k^3$ is the integration volume in Fourier space for the bispectrum. Notice that for the Gaussian term to be non-zero the triangles $a$ and $b$ are identical; that is why we wrote the subscript $aa$ for  $\mathcal{M}^{(\alpha\beta\kappa\lambda)}_{\mu\nu,aa}$.

Up to order $\propto\delta^8$ in terms of the matter over-density perturbation,  there are only two non-Gaussian terms contributing to the bispectrum covariance (see eq. \ref{eq:bk_cov_expansion}) since the third one, proportional to the pentaspectrum, is of order $\propto\delta^{10}$ \cite{Bernardeau:2001qr,Sefusatti:2006pa}. The first is proportional to the product of two bispectra:

\begin{align}
\label{eq:cov_bb_ngbb}
&\mathrm{C}_{\propto\mathrm{BB}}^{\mathrm{B}^{\alpha\beta}_{\mathrm{g}}\mathrm{B}^{\kappa\lambda}_{\mathrm{g}}}\left(k_1,k_2,k_3;k_4,k_5,k_6\right) = 
\dfrac{(2\pi)^3\mathcal{C}^{\alpha\beta}_{\kappa\lambda}\delta^{\mathrm{K}}_{34} }{4\pi k_3k_4\Delta k V_\mathrm{s}}
\notag\\
&\times \dfrac{1}{8}
\int_{-1}^{+1}d\mu_3d\mu_1d\mu_5 
\mathcal{M}^{(\alpha\beta\kappa\lambda)}_{\mu\nu,ab}
\mathrm{B}_{\mathrm{g}}^{\mathrm{s}}\left(k_1,k_2,k_3,\mu_3,\mu_1\right)\mathrm{B}_{\mathrm{g}}^{\mathrm{s}}\left(k_3,k_5,k_6,\mu_3,\mu_5\right) \; + 8 \;\mathrm{perm.}
,
\end{align}

\noindent for which in total there are nine possible permutations. The second term is proportional to the product of the power spectrum and trispectrum:

\begin{align}
\label{eq:cov_bb_ngpt}
&\mathrm{C}_{\propto\mathrm{TP}}^{\mathrm{B}^{\alpha\beta}_{\mathrm{g}}\mathrm{B}^{\kappa\lambda}_{\mathrm{g}}}\left(k_1,k_2,k_3;k_4,k_5,k_6\right) = 
\dfrac{\delta^{\mathrm{K}}_{34}(2\pi)^3\mathcal{C}^{\alpha\beta}_{\kappa\lambda}\Delta D}{4\pi k_3k_4\Delta k_3\Delta k_4V_\mathrm{s}}
\notag \\
&\times\dfrac{1}{8}
\int^{+1}_{-1}d\mu_Dd\mu_1d\mu_5
\mathrm{T}_{\mathrm{g}}\left(k_1,k_2,k_5,k_6,D,\mu_D,\mu_1,\mu_5\right)\mathrm{P}_{\mathrm{g}}\left(D,-\mu_D\right)
\; + 8 \;\mathrm{perm.}
\,,
\end{align}

\noindent where $k_3 = k_4 = D$ is the diagonal of the quadrilateral configuration defining the trispectrum. Before the angular integration, eight coordinates are needed in order to describe a galaxy trispectrum configuration. Differently from what we had for the previous non-Gaussian term in Equation \eqref{eq:cov_bb_ngbb}, $\delta_D(\mathbf{q}_3-\mathbf{q}_4)$, in this case the Kronecker's delta condition derives from $\delta_D(\mathbf{q}_3+\mathbf{q}_4)$ and that is the reason for having $-\mu_D$ as second argument of the galaxy power spectrum in the integrand (even if it actually does not make any difference at tree-level). More detailed explanation about the trispectrum are reported in Appendix \ref{sec:trispectrum}.

Finally we model at lowest order the cross-correlation between the galaxy power spectrum and bispectrum multipoles by computing (for the five-point correlator originating this term see Equation \ref{eq:pkbk_cov_expansion}):

\begin{align}
\label{eq:cross_pkbk}
\mathrm{C}_{\propto\mathrm{PB}}^{\mathrm{P}^{\ell}_{\mathrm{g}}\mathrm{B}^{\alpha\beta}_\mathrm{g}}\left(k_1;k_2,k_3,k_4\right) &=
\dfrac{2\,(2\pi)^3\mathcal{C}_{\alpha\beta}^{\ell}\delta^{\mathrm{K}}_{12}}{4\pi k_2^2 \Delta kV_\mathrm{s}}
\notag \\
&\times\dfrac{1}{4\pi}\int^{+1}_{-1}\int^{2\pi}_{0}d\phi\mathcal{M}^{(\ell\alpha\beta)}_{\mu_a\mu_b\nu_b}
\mathrm{P}_{\mathrm{g}}^{\mathrm{s}}\left(k_2,\mu_2\right)
\mathrm{B}_{\mathrm{g}}^{\mathrm{s}}\left(k_2,k_3,k_4,\mu_2,\phi\right) \; + 2 \;\mathrm{perm.}\,.
\end{align}{}
 The $\Delta k$ appearing at the denominator corresponds to the  Fourier space  resolution adopted for the power spectrum  (i.e., giving the number of  power spectrum modes available in Fourier space). The bispectrum has its own $\Delta k$ which does not have to coincide with the one chosen for the power spectrum, but does not appear here\footnote{As shown in Equation  \ref{eq:cross_cov_pkbk} the bispectrum  $\Delta k$ cancels out.}. For the power spectrum, the multipole expansion is done in terms of $\mu_a$, the cosine of the angle between the $k$-vector and the line of sight. The angle $\phi$ is the same as the one defined below Equation \eqref{eq:bk_model}.

\begin{figure}[tbp]
\centering 
\includegraphics[width=.98\textwidth]
{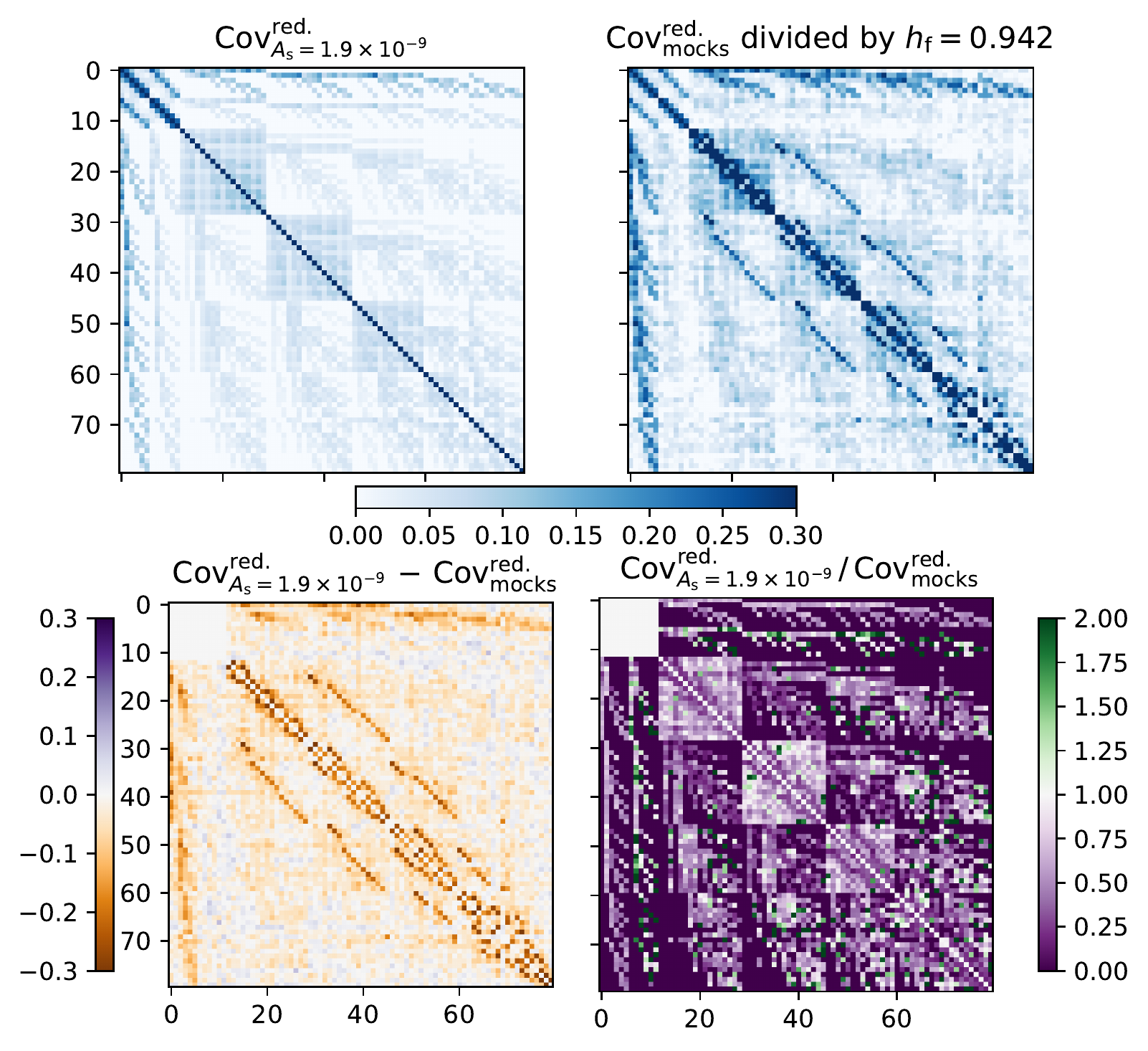}
\caption{\label{fig:covariance_comparison2D} Reduced analytical (top-left) and numerical (top-right) covariance matrices for the $\mathrm{\left[\mathrm{P}_\mathrm{g}^{(0,2)};\mathrm{B}_\mathrm{g}^{(00)}\right]}$ data vector, $\Delta k_5$ case. In the bottom row are also shown the difference (left) and ratio (right) between the two covariance matrices. In order to minimise the potential systematics given by analytically model the covariance matrix, we used the auto-covariance of the power spectrum multipoles estimated from the mocks also in our analytical (hybrid) covariance. That is the reason why in the above Figure there is no difference between analytical and numerical covariances for $\mathrm{P}_\mathrm{g}^{(0,2)}$. 
From the bottom row we can see that the analytical model misses certain features close to the diagonal for the bispectrum auto-covariance.
In the cross correlation part, the cross correlation between power spectrum monopole and bispectrum monopole is slightly underestimated while the one with the power spectrum quadrupole is slightly overestimated.}
\end{figure}

\section{Numerical and hybrid covariance matrix for \texorpdfstring{$\mathbf{\left[{P}_{g}^{(0,2)};{B}_{g}^{(00)}\right]}$}{}}
\label{sec:compare_covs}

We compare the analytical covariance matrix for the data vector $\mathrm{\left[\mathrm{P}_\mathrm{g}^{(0,2)};\mathrm{B}_\mathrm{g}^{(00)}\right]}$ with a numerical one derived by measuring the above statistics from 1400 realisations of the MultiDark Patchy BOSS DR12 galaxy catalogues \cite{Kitaura:2015uqa,Rodriguez-Torres:2015vqa}. The Patchy mocks reproduce the number density, selection function, survey geometry and clustering properties of BOSS DR12 galaxy sample \cite{Alam:2016hwk}; for each mock the power spectrum monopole and quadrupole are available and we have calculated the bispectrum monopole for both North and South Galactic Caps (NGC and SGC).

The same fiducial cosmology of the mocks, $\Omega_{\Lambda} = 0.693$, $\Omega_{\mathrm{m}}(z=0) = 0.307$, $\Omega_{\mathrm{b}}(z=0) = 0.048$, $\sigma_8(z=0)=0.829$, $n_{\mathrm{s}} = 0.96$, $h_0 = 0.678$, has been used the generate the necessary analytical matter power spectrum \cite{Lesgourgues:2011re}  for our modeling.

Note that the covariance matrix obtained from the  survey mocks has a contribution due to the survey  window function. The analytic expression for the covariance matrix does not account for the  extra mode-coupling and thus coupling between different statistics, induced by the window. 
In  our analytically-generated data vector used for  parameter  forecast  we do not include the survey window effect, hence we do not worry too much about modeling the same effect in the covariance matrix.  More discussion on the effect of the window can be found below and in Appendix \ref{sec:window_effect}.
In order to model the window additional contribution to statistical error one could proceed as in \cite{Hinshaw:2003ex} for the CMB angular power spectrum: under the assumption of the mask correction being small with respect to the signal covariance, the total covariance could be written as a sum $\mathrm{C_{tot}} = \mathrm{C_{signal}} + \mathrm{C_{mask}}$.

Similarly to  \cite{Sugiyama:2019ike} we set the bias parameters values $b_1 = 2.0$ and $b_2=0.$, together with using the same effective survey volume $V_\mathrm{s} = 1.76 \left[h^{-1}\,\mathrm{Gpc}\right]^3$. The choice for the linear bias $b_1$ is not too dissimilar from what obtained in the BOSS clustering analysis \cite{Gil-Marin:2016wya} while the survey volume derived in \cite{Sugiyama:2019ike} is not too far from the one ($V_\mathrm{s} = 1.92 \left[h^{-1}\,\mathrm{Gpc}\right]^3$) fitted from comparing the numerical covariance estimated from the mocks and its theoretical model in the 3-pt correlation function analysis done in \cite{Slepian:2015hca}.

\begin{figure}[tbp]
\centering 
\includegraphics[width=.90\textwidth]
{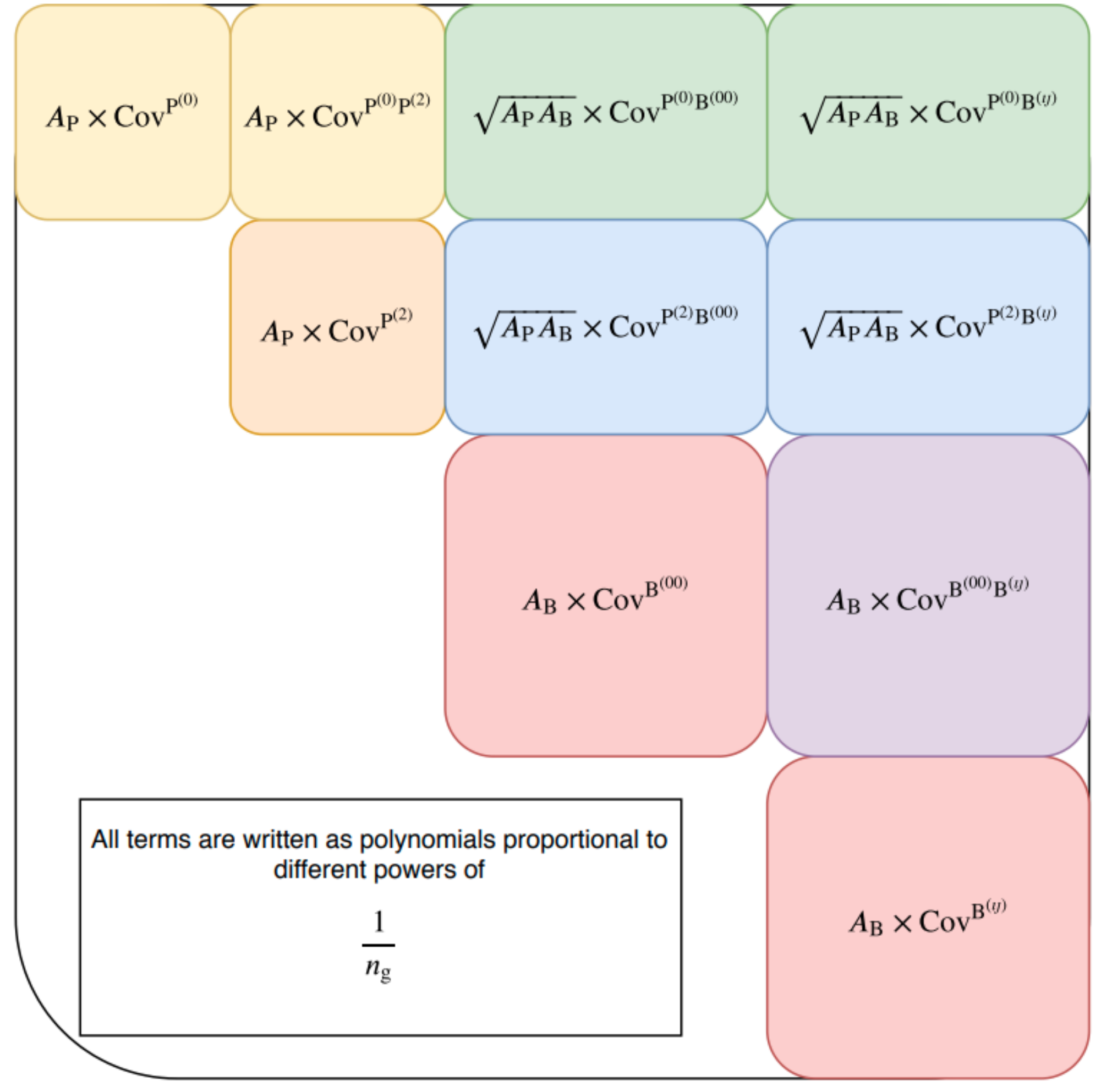}
\caption{\label{fig:calibrate_covariance}Analytical covariance matrix: calibrated terms. The powers of the average galaxy number density enter the different components of the data vector's covariance as shown in Equation \ref{eq:sn_expansion}. From this figure we can see the difference between fitting the two power spectrum and bispectrum covariance amplitude parameters $A_\mathrm{P}$, $A_\mathrm{B}$ and the physical scalar perturbation amplitude parameter $A_\mathrm{s}$. The power spectrum part of the covariance is proportional to $A_\mathrm{s}^2$ (Gaussian term, Equation \ref{eq:cov_pp} ), the bispectrum part to both $A_\mathrm{s}^3$ (Gaussian term, Equation \ref{eq:cov_bp_gg}) and to $A_\mathrm{s}^4$ (non-Gaussian term, Equations \ref{eq:cov_bb_ngbb} and \ref{eq:cov_bb_ngpt}); the cross-covariance terms are proportional to $A_\mathrm{s}^3$ (Equation \ref{eq:cross_pkbk}). To calibrate $A_\mathrm{s}$ would then imply changing not only the balance between different parts of the covariance but also, as for the case of the bispectrum, the balance between Gaussian and non-Gaussian contributions belonging to the same terms. This is the reason why we prefer to estimate a priori through an MCMC sampling an effective value of $A_\mathrm{s}$. Only afterwards, after the analytical covariance expressions have been computed, we calibrate the two parameters $A_\mathrm{P}$, $A_\mathrm{B}$. These can be interpreted as corrections to the analytically computed Fourier integration volume for each covariance matrix component.}
\end{figure}

In the power spectrum multipoles covariance matrix expression we set $\Delta k = 0.01 h/$Mpc, which is the same value used in numerical measurement pipeline. For the galaxy bispectrum we use two different $\Delta k$ bin sizes proportional to the fundamental frequency $k_f = \frac{(2\pi)^3}{V_\mathrm{b}}$ where $V_\mathrm{b} = (3500\,\mathrm{Mpc}/h)^3$ is the survey volume for the mock catalogues cubic box. We select the two cases: $\Delta k_{6,5} = 6,5 \times k_f$ which given our algorithm \cite{Gil-Marin:2016wya} to generate $(k_1,k_2,k_3)$ triplets imposing the triangle closure condition and the range of scales considered,  corresponds to using 32 and 68 triangle configurations, respectively. In the main body of this work all the results for the $\Delta k_5$ case have been presented. The key figures and table for the $\Delta k_6$ case are shown in Appendix \ref{sec:deltak6}.

We adopt a very conservative $k$-range for the sides of the triangle configurations used for the bispectrum multipoles, in particular we set $k_\mathrm{min} = 0.03\, h$/Mpc and $k_\mathrm{max} = 0.09\, h$/Mpc. The reason for choosing a low $k_\mathrm{max}$ is twofold. Firstly, this allows us to trust our standard perturbation theory calculations at tree-level, for example for what concerns the cross-correlation between power spectrum and bispectrum as noted in \cite{Sugiyama:2019ike}. Secondly, this choice limits the number of triangle configurations  in the data vector. This is essential in order to make it feasible to analytically compute in a reasonable amount of time all the terms of the covariance matrix with a sufficient precision. Moreover, the longer the data vector, the less numerically stable is the inversion of its covariance matrix needed for the likelihood evaluation.  As already pointed  out in  \cite{Gualdi:2019sfc} this is the case when several multipoles of the bispectrum are present simultaneously in the data vector without compression.

Nevertheless by choosing $k_\mathrm{max} = 0.09\, h$/Mpc our analysis uses only a fraction of the baryonic acoustic oscillations  information encoded in  the galaxy density field. To fully capture and exploit this information and in particular its anisotropic component using the bispectrum multipoles, it would be necessary to further improve the covariance analytical model to extend its reliability to higher $k$'s. For this reason our finding might be conservative.

In order to choose the scalar amplitude parameter $A_\mathrm{s}$ needed to compute the linear matter power spectrum (with CLASS \cite{Lesgourgues:2011re}) to be used in the analytical covariance matrix template, we run a preliminary MCMC sampling using the numerical covariance matrix and the mean of the mocks as data vector. We then leave  $A_\mathrm{s}$ free to vary together with the additional parameters $(b_1,b_2,f,\alpha_\perp, \alpha_\parallel)$. We obtain $A_\mathrm{s}\simeq 1.9 \times 10^{-9}$ as a best-fit value, we recognize that this is  lower than the value reported in the Planck analysis \cite{Aghanim:2018eyx} ($A_\mathrm{s}=3.04\times10^{-9}$). This is just a shortcut to make our calculations numerically stable. In fact, as reported below, and explained in Fig.~\ref{fig:calibrate_covariance} the effective amplitude of the power spectrum and bispectrum  covariances are later free parameters of the normalisation.

For the covariance computation we set the FoG parameters in Equation \ref{eq:fog_parameters} for the power spectrum and bispectrum multipoles to the best fit values obtained in the BOSS analysis \cite{Gil-Marin:2016wya}, $\sigma_\mathrm{P}=3.5$ and $\sigma_\mathrm{B}=7.39$. In fact, even if we limit our $k$-range to linear scales where the FoG effect should be negligible, in  some of the covariance matrix  terms (in particular, shot-noise correction terms)  the sums of two or three $k$-vectors  appear in the argument of the power spectrum or bispectrum.  This is not the case for the terms proportional to the trispectrum and hence no FoG parameter is  used for the trispectrum. In the data vector's model used for the MCMC samplings both $\sigma_\mathrm{P}$ and $\sigma_\mathrm{B}$ have been set to zero since we limit our analysis to linear scales. 

To calibrate the covariance matrix we adjust three parameters:
 the average number density of galaxies $n_\mathrm{g}$, an effective amplitude for the power spectrum part of the covariance $A_{\mathrm{P}}$ and another effective amplitude $A_{\mathrm{B}}$ for the bispectrum part. These parameters act on the calibrated terms as shown in Figure \ref{fig:calibrate_covariance}. Figure \ref{fig:calibrate_covariance} clarifies the difference between fitting the two amplitude parameters $A_\mathrm{P}$, $A_\mathrm{B}$ and the physical scalar amplitude $A_\mathrm{s}$. While $A_\mathrm{s}$ is common to all the terms in the covariance, each is term proportional to a different power of the scalar perturbations amplitude parameter (see caption of Figure \ref{fig:calibrate_covariance} for details), $A_\mathrm{P}$ and $A_\mathrm{B}$ can be interpreted as separate corrections to the analytically-computed integration volumes in Fourier space for power spectrum ($V_\mathrm{p} = 4\pi\Delta k k^2 $) and bispectrum ($V^{\mathrm{t}}_{123} = 8\pi^2k_1k_2k_3\Delta k_1\Delta k_2\Delta k_3$). 
 The analytical expressions for these volumes, which are used to normalise the statistics estimators by the number of modes measured (see Appendix A of \cite{Gualdi:2018pyw} for a detailed explanation), are obtained in the "infinitesimal" bin-width approximation. 
In particular using $\Delta_k=5,6\times k_\mathrm{f}$ and limiting our analysis to small $k$'s the condition $k\ll\Delta_k$ may not be optimal. Moreover, $A_\mathrm{P}$, $A_\mathrm{B}$ can also account for a mismatch between the fiducial effective survey volume value and the real one. Therefore it is not surprising that a calibration of these quantities is needed when they are not measured but computed analytically; reassuringly, their value is of the order of unity as reported in Table \ref{tab:fit_cov_params}.

 For what concerns $n_\mathrm{g}$, we treat it as a free parameter in order to model shot-noise deviations from  Poissonian behaviour. This is non uncommon and often also used when analyzing galaxy surveys. Indeed the expressions that we use to account for the shot-noise terms in the covariance model were derived under the Poissonian approximation (see Ref.~\cite{Sugiyama:2019ike}). Even letting $n_\mathrm{g}$ free to vary, we  find a best fit value very similar to the one computed  in Ref.~\cite{Sugiyama:2019ike} for the same data-set.

 We fit these parameters by comparing the diagonal of the analytical covariance matrix with the one numerically estimated using a set of 1400 Patchy galaxy mock catalogue realisations after correction by the  Hartlap factor $h_\mathrm{f} = (N_\mathrm{mocks} - N_\mathrm{dv} -2)/(N_\mathrm{mocks}-1)$ \cite{Hartlap:2006kj}, where $N_\mathrm{dv}$ is the data vector's size while $N_\mathrm{mocks}$ is the number of galaxy catalogues used to estimate the covariance (1400 in our case). 
In summary calibration is done following the expression below:

\begin{eqnarray}
\label{eq:sn_expansion}
C_\mathrm{tot}^{\mathrm{P}^{(\ell)},\mathrm{B}^{(\imath,\jmath)}} = \begin{cases}

A_{\mathrm{P}} \times \left[C_{\mathrm{P}^{\ell}}^{n_\mathrm{g}^0} + \dfrac{1}{n_\mathrm{g}}C_{\mathrm{P}^{\ell}}^{n_\mathrm{g}^{-1}}+\dfrac{1}{n_\mathrm{g}^2}C_{\mathrm{P}^{\ell}}^{n_\mathrm{g}^{-2}}\right]
, & \mbox{for } \mathrm{P}^{(\ell)}_\mathrm{g} \mbox{ cov.}  \\ 
A_{\mathrm{B}} \times \left[C_{\mathrm{B}^{\imath\jmath}}^{n_\mathrm{g}^0} + \dfrac{1}{n_\mathrm{g}}C_{\mathrm{B}^{\imath\jmath}}^{n_\mathrm{g}^{-1}}
+\dfrac{1}{n_\mathrm{g}^2}C_{\mathrm{B}^{\imath\jmath}}^{n_\mathrm{g}^{-2}}
+\dfrac{1}{n_\mathrm{g}^3}C_{\mathrm{B}^{\imath\jmath}}^{n_\mathrm{g}^{-3}}
+\dfrac{1}{n_\mathrm{g}^4}C_{\mathrm{B}^{\imath\jmath}}^{n_\mathrm{g}^{-4}}
\right]
, & \mbox{for } \mathrm{B}^{(\imath\jmath)}_\mathrm{g} \mbox{ cov.} \\
\sqrt{A_{\mathrm{P}}A_{\mathrm{B}}} \times \left[C_{\mathrm{P}^{(\ell)}_\mathrm{g}\mathrm{B}^{\imath\jmath}}^{n_\mathrm{g}^0} + \dfrac{1}{n_\mathrm{g}}C_{\mathrm{P}^{(\ell)}_\mathrm{g}\mathrm{B}^{\imath\jmath}}^{n_\mathrm{g}^{-1}}
+\dfrac{1}{n_\mathrm{g}^2}C_{\mathrm{P}^{(\ell)}_\mathrm{g}\mathrm{B}^{\imath\jmath}}^{n_\mathrm{g}^{-2}}
+\dfrac{1}{n_\mathrm{g}^3}C_{\mathrm{P}^{(\ell)}_\mathrm{g}\mathrm{B}^{\imath\jmath}}^{n_\mathrm{g}^{-3}}
\right]
, & \mbox{for } \mathrm{P}^{(\ell)}_\mathrm{g}\mathrm{B}^{(\imath\jmath)}_\mathrm{g} \mbox{ cross-cov.}
\end{cases}
\end{eqnarray}{}

\noindent In Table \ref{tab:fit_cov_params} are reported the values for the fitted parameters both in the $\Delta k_5$ and $\Delta k_6$ cases.

\renewcommand{\arraystretch}{1.5}
\begin{table}[tbp]
\centering
\begin{tabular}{|c|c|ccc|cc|}
\hline
$k$-bin size & $N_\mathrm{tr.}$ &$A_\mathrm{P}$& $A_\mathrm{B}$ & $n_\mathrm{g}\;\left[h/\mathrm{Mpc}\right]^3$ & $\mathrm{R_c}=\langle \mathrm{Cov_{ana.}^{ii}}/\mathrm{Cov_{num.}^{ii}} \rangle$ & $\sigma \left(\mathrm{R_c}\right)$ \\
\hline
$\Delta k_6$ & 32  & 1.457 & 1.024 & 3.49 $\times 10^{-4}$ & 0.997 & 0.057  \\
$\Delta k_5$ & 68  & 1.480 & 0.868 & 3.25 $\times 10^{-4}$ & 0.998  & 0.047 \\
\hline
\end{tabular}
\caption{\label{tab:fit_cov_params} Values for the analytic covariance matrix parameters fitted by comparing the diagonal  elements with  those  obtained  numerically from the mock catalogues. On the last two columns are shown the average and the standard deviation relative to the ratio between the analytical and numerical covariance matrices diagonals. The fitted values of $n_\mathrm{g}$ are very close, especially in the $\Delta k_5$ case, to the one computed in \cite{Sugiyama:2019ike}.}
\end{table}

\begin{figure}[tbp]
\centering 
\includegraphics[width=.98\textwidth]
{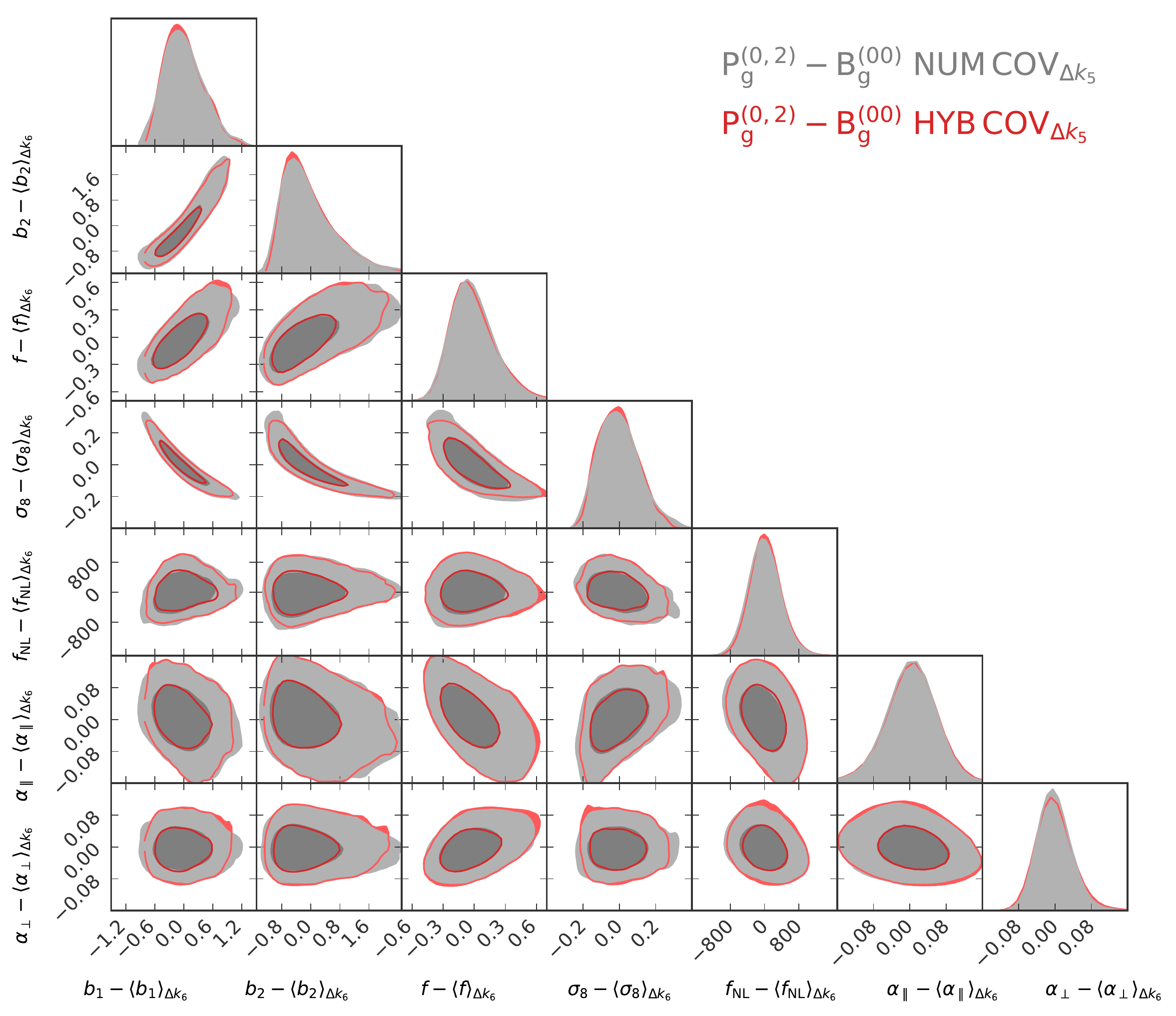}
\caption{\label{fig:p02b00p6dk5}
1 and 2D marginalised posterior distributions for the data vector $\mathrm{\left[\mathrm{P}_\mathrm{g}^{(0,2)};\mathrm{B}_\mathrm{g}^{(00)}\right]}$, $\Delta k_5$ case. In gray (NUM COV) are shown the posterior contours given by employing the numerical covariance matrix estimated from the mocks to evaluate the likelihood, while in red (HYB COV) the equivalent is shown for the case in which the analytic/hybrid covariance has been used for the same purpose. Both 1-2D,  68$\%$ and 95$\%$ confidence intervals show good agreement with no significant difference in terms of the shape of the degeneracies present between the different parameters.}
\end{figure}

\subsection{Visual comparison }
We begin by visually inspecting the differences between the analytical covariance matrix (after calibration) and the  numerical.  Figure \ref{fig:covariance_comparison1D}  shows  1D sections (columns/rows) for both the bispectrum autocovariance and its cross-correlation with the power spectrum for the $\Delta k_5$ case.
The shaded area is obtained by splitting the 1400 mocks into ten groups and then computing the numerical covariance matrix for each of them. From the ten estimated covariance matrices we compute each element's standard deviation as a proxy for the statistical noise. This gives us a possible estimate of the level of agreement between analytical and numerical covariance matrices.
The bispectrum auto-covariance results are well approximated (mostly within the shaded area) by the analytical model. 

The cross-covariance between bispectrum and power spectrum multipoles shows a larger discrepancy between numerical and analytical covariance matrices. In particular the amplitude of certain elements seems to be lower in the analytical case. For other "pixels" the analytical value is exactly equal to zero due to the Kronecker's delta condition from Equation \ref{eq:cross_pkbk}. The absence in our model of higher order terms such as the one proportional to the tetraspectrum could explain the lower amplitude observed in the analytical lines in the right panel of Figure \ref{fig:covariance_comparison1D}. In the pixels where the analytical covariance matrix is by definition equal to zero we see that the shaded area indicates that also the numerical estimate is compatible with zero and therefore the non-null values seen in the numerical covariance matrix could be interpreted simply as statistical noise. Another source of discrepancy could be the lack of the survey window selection effect in our covariance template (more below).
Finally the shot-noise correction appears to be relatively negligible with respect to the non-Gaussian terms for the off-diagonal elements.

It also results useful to look at the whole 2D analytical and numerical reduced covariance matrices (i.e., element of row i and column j is divided by the square root of the product of element ii times element jj). In Figure \ref{fig:covariance_comparison2D} are shown the reduced covariance matrices for the $\Delta k_5$ binning case\footnote{Another reason to choose this binning case for the whole analysis is that there are enough triangle configurations to resolve the structure and features present in the covariance matrix but not too many to make them invisible due to the limited size of the image.}. The comparison between the numerical and analytical covariance reveals qualitative good agreement. The only relevant difference is the lack, in the calibrated hybrid template, of the features parallel to the covariance diagonal. 
These features are due to the additional mode-coupling induced by the survey window.  In Appendix \ref{sec:window_effect} we compare our hybrid model to the numerical covariance obtained from a set of 440 simulated galaxy fields in cubic boxes with periodic boundary conditions \cite{Chuang:2014vfa}. In Figure \ref{fig:covariance_comparison2D periodic} we see that in the absence of survey mask and window, such features are not present in  the covariance matrix. It will become evident in Section \ref{sec:5.2} that these features have no significant effect on our analysis.

\begin{figure}[tbp]
\centering 
\includegraphics[width=.98\textwidth]
{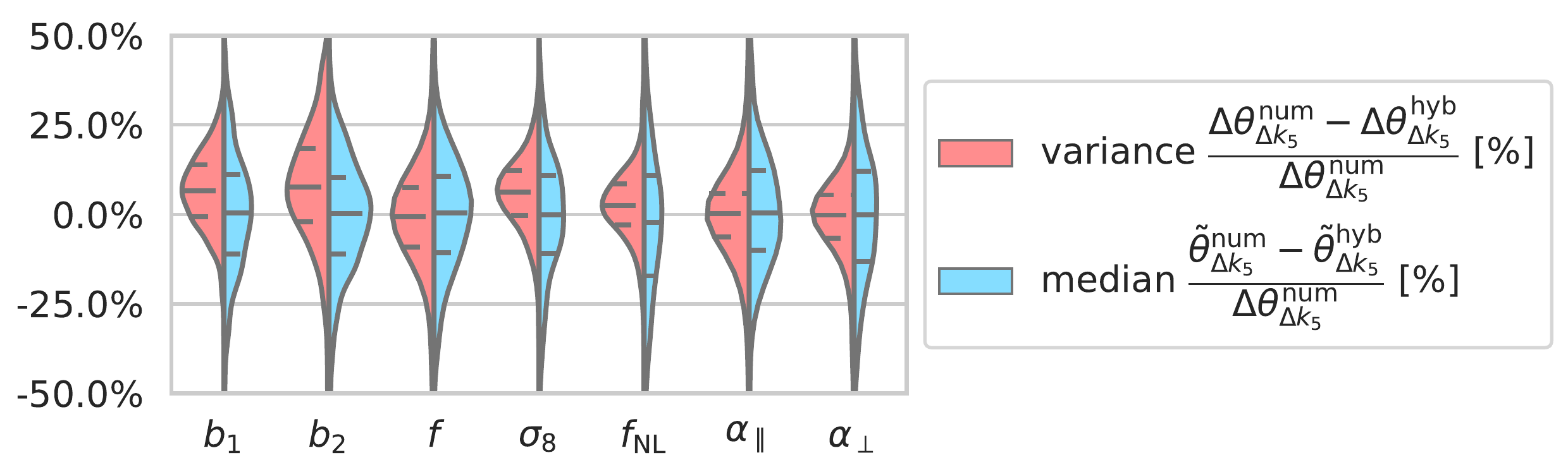}
\caption{\label{fig:stat_test_p02b00_violin} 
Relative difference between analytical and numerical covariance matrix results for what concerns the 1D 68$\%$ confidence intervals and median values obtained from running a MCMC sampling using as data vector each measurement of $\mathrm{\left[\mathrm{P}_\mathrm{g}^{(0,2)};\mathrm{B}_\mathrm{g}^{(00)}\right]}$ from the 1400 mock catalogues. 
The horizontal lines inside the violins delimitate the central quartiles of the ratios distribution.
From the violin plots above it is possible to infer that, for the vast majority of the mocks, the difference between the values obtained using the analytical and numerical covariance matrices is less than $25\%$ the value of the 1D 68$\%$ confidence intervals obtained using the numerical covariance matrix.}
\end{figure}

\begin{figure}[tbp]
\centering 
\includegraphics[width=.98\textwidth]
{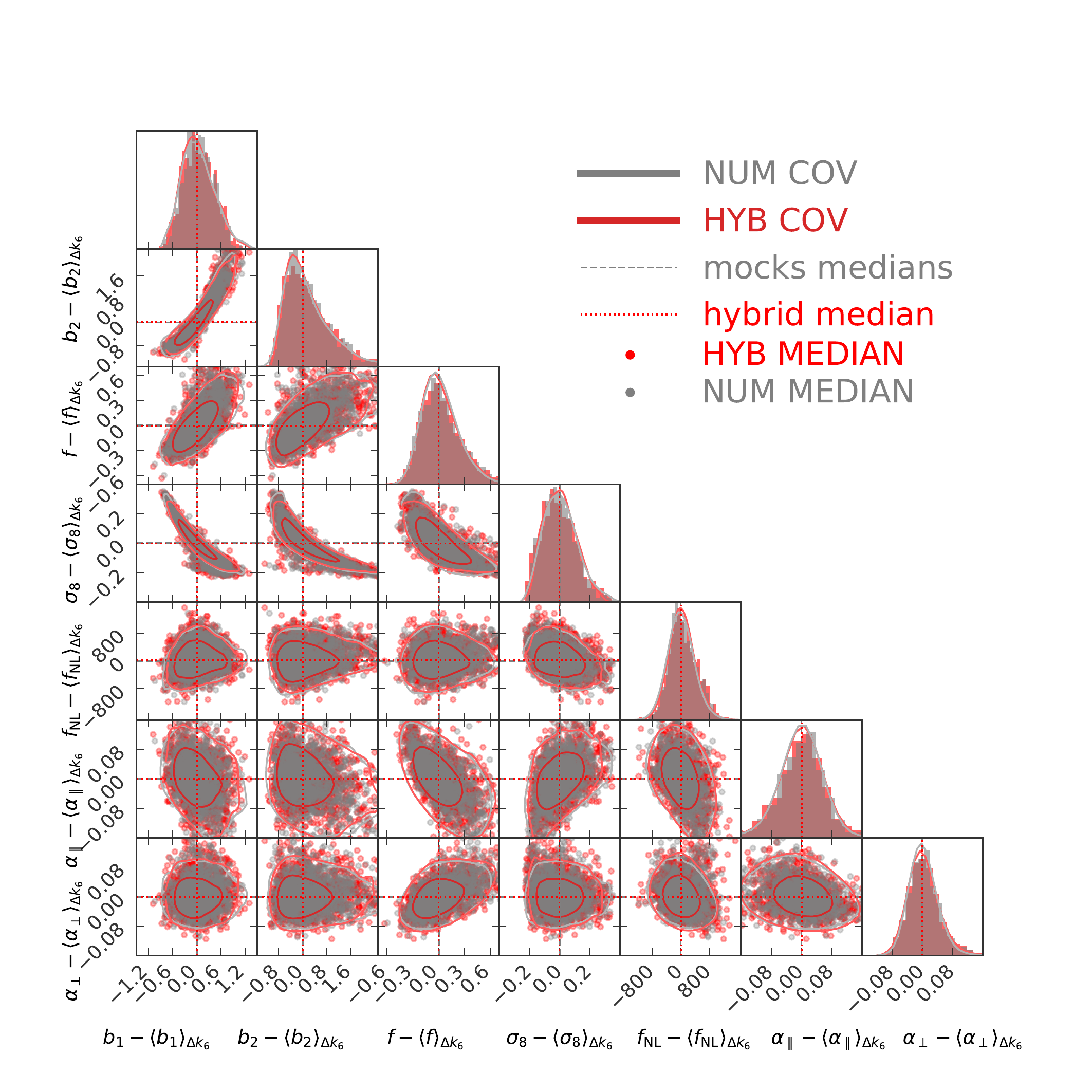}
\caption{\label{fig:stat_test_p02b00_bfdistribution} Cosmological parameters posterior distributions for  $\mathrm{\left[\mathrm{P}_\mathrm{g}^{(0,2)};\mathrm{B}_\mathrm{g}^{(00)}\right]}$ data vector, $\Delta k_5$ case. In gray (MOCKS COV) and red (HYB COV)  are shown the 1-2D marginalised posterior distributions for using as data vector the mean of the mocks together with the numerical and hybrid covariance matrices to evaluate the likelihood, respectively. On top of these contours we show the scatter of the median  values for all the model parameters obtained for each of the 1400 galaxy catalogues, again using both numerical and hybrid covariance matrices. 
The similarity of the posteriors and the scatter among the mocks adds further confidence in the reliability of our fitted analytical model of the data vector's covariance.}
\end{figure}

\subsection{Constraints comparison for the (reduced)  data vector 
\texorpdfstring{$\mathbf{\left[ P_g^{(0,2)};B_g^{(00)}\right]}$}{}}
\label{sec:5.2}
We estimate the  cosmological posterior for the (reduced) data vector $\mathbf{\left[ P_g^{(0,2)};B_g^{(00)}\right]}$ using the analytical/hybrid covariance matrix and the numerical one.
Whenever we compare the two covariance matrices we use for each test the same data vector. In this section we either use the mean measurement from the mocks or each individual mock measurement. 

To compare the two posteriors so obtained, we consider a seven parameters set:\\   $(b_1,b_2,f,\sigma_8,f_\mathrm{NL},\alpha_\perp, \alpha_\parallel)$, where
$b_1$ and $b_2$ are the first and second order Eulerian bias parameters, $f$ is the linear growth-rate, $\sigma_8$ is the normalisation of the matter power spectrum, $f_\mathrm{NL}$ is the amplitude of local primordial non-Gaussianity perturbations \cite{Verde:1999ij,Komatsu:2001rj,Bartolo:2004if,Byrnes:2010em} and $\alpha_\perp$, $\alpha_\parallel$ are the Alcock-Paczy\`{n}sky parameters. 

Following what is done in the BOSS analyses \cite{Gil-Marin:2014sta,Gil-Marin:2016wya} we use the local Lagrangian bias approximation to fix the value of the non-local tidal bias parameter $b_{\mathrm{s}_2}$ appearing in the second order kernel of the bispectrum to $b_{\mathrm{s}_2}=-4(b_1-1)/7$. We have checked through a specific MCMC sampling that allowing this parameter to vary does not significantly change our results.

This is shown in Figure  \ref{fig:p02b00p6dk5} for the $\Delta k_5$ case (for which there are 68 triangle configurations in the bispectrum data vector). 

Since we focus on the change in the parameter constraints width, we show our results with each parameter's central value (median) subtracted. The central value is given by the analysis performed using the numerical covariance matrix. A full parameters constraints analysis will be presented elsewhere.

From Figure \ref{fig:p02b00p6dk5} we can see that qualitatively there is very good agreement between the posterior distributions obtained with the hybrid and numerical covariance matrices. 
With another separate MCMC run we tested what happens in both cases:  setting to zero all the off-diagonal terms or only the bispectrum auto-covariance ones. The resulting marginalised 1-2D posterior distributions are- not unexpetedly- significantly tighter then the ones obtained using the covariance matrix estimated from the mocks.

We go even further to quantify the performance of the hybrid covariance matrix approach: we perform a MCMC sampling on the data vector measurement from each of 1400 mock catalogues using both analytic and the hybrid covariance matrices.

We then compare in Figure \ref{fig:stat_test_p02b00_violin} the probability distributions 1D 68$\%$ confidence intervals and the medians for each model parameter. The relative difference between variances and medians obtained using numerical and hybrid covariances is for the large majority of the mocks less than $\sim 25\%$ the value of each parameter's variance.
In Figure \ref{fig:stat_test_p02b00_bfdistribution} we show the 1 and 2D posterior distributions for the parameters and the  scatter of the median\footnote{As it is common to do in Bayesian inference, we stick to the so-called Central Credible Interval which reports  errors around the median of a distribution. Posteriors are very symmetric hence the median is very close to the mean and the maximum anyway.} values for each mock realisation.  Point and lines in red correspond to the analytic/hybrid covariance, points and lines in gray correspond to the numerical covariance.

From these quantitative and statistical tests we conclude that our analytical model, once  calibrated (i.e., fitting the three parameters $A_\mathrm{P},\,A_\mathrm{B},\,n_\mathrm{g}$), represents a realistic proxy for the covariance matrix estimated from the galaxy mock catalogues for the $\mathrm{\left[\mathrm{P}_\mathrm{g}^{(0,2)};\mathrm{B}_\mathrm{g}^{(00)}\right]}$ data vector.

Given this result, the main assumption of this work is to consider reliable the extension of our covariance model to the galaxy bispectrum higher multipoles. This is motivated by the fact that  our choice of fitting parameters does not depend on the multipole chosen but just on the statistic (power spectrum or bispectrum).

\section{Higher multipoles of the bispectrum:
\texorpdfstring{$\mathbf{B_g^{(20)}, B_g^{(02)}, B_g^{(40)}, B_g^{(04)}, B_g^{(22)}}$}{}}
\label{sec:higher_bk_multi}

\begin{figure}[tbp]
\centering 
\includegraphics[width=.98\textwidth]
{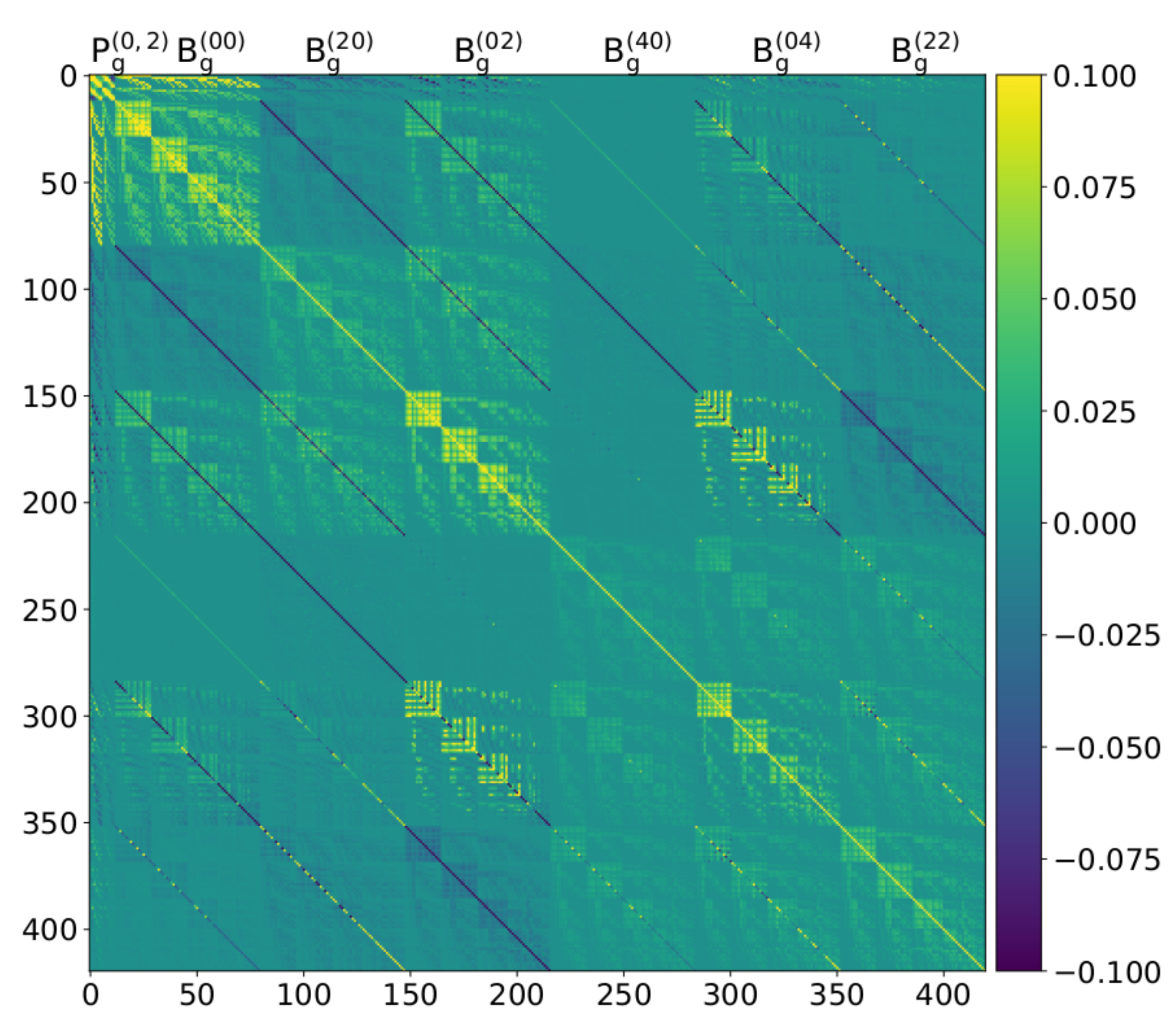}
\caption{\label{fig:red_cov_p02b002002_6dk}  Reduced covariance matrix for the data vector including the galaxy bispectrum higher multipoles,  $\mathrm{\left[\mathrm{P}_\mathrm{g}^{(0,2)};\mathrm{B}_\mathrm{g}^{(00,20,02,40,04,22)}\right]}$, for the $\Delta k_5$ case. It is interesting to notice the different sign of the correlation between the bispectrum multipoles. For example $\mathrm{B}_\mathrm{g}^{(00)}$ and $\mathrm{B}_\mathrm{g}^{(20)}$ are anti-correlated in the off-diagonal terms whilst $\mathrm{B}_\mathrm{g}^{(00)}$ and $\mathrm{B}_\mathrm{g}^{(02)}$ are positively correlated. This information hints to the additional constraining power achievable through the employment of both $\mathrm{B}_\mathrm{g}^{(20)}$ and $\mathrm{B}_\mathrm{g}^{(02)}$ since they are differently correlated to the isotropic signal, $\mathrm{B}_\mathrm{g}^{(00)}$.}
\end{figure}

We now proceed to estimate the added value given by extending the clustering statistics data vector to the galaxy bispectrum higher multipoles.  We use the covariance analytical model presented in Section \ref{sec:cov_matrix_model} to sample the model parameters posterior distribution. In Figure \ref{fig:red_cov_p02b002002_6dk} the resulting reduced covariance matrix for the data vector $\left[\mathrm{P_g^{(0,2)}};B^{(00,20,02,40,04,22)}_g\right]$ is displayed. As an example it is interesting to notice that bispectrum monopole $\mathrm{B_g^{(00)}}$ is negatively correlated with $\mathrm{B_g^{(20)}}$ while positively correlated with $\mathrm{B_g^{(02)}}$. 
The difference in the correlations between different bispectrum multipoles terms suggests that the multipoles have complementary constraining power.

\begin{figure}[tbp]
\centering 
\includegraphics[width=.98\textwidth]
{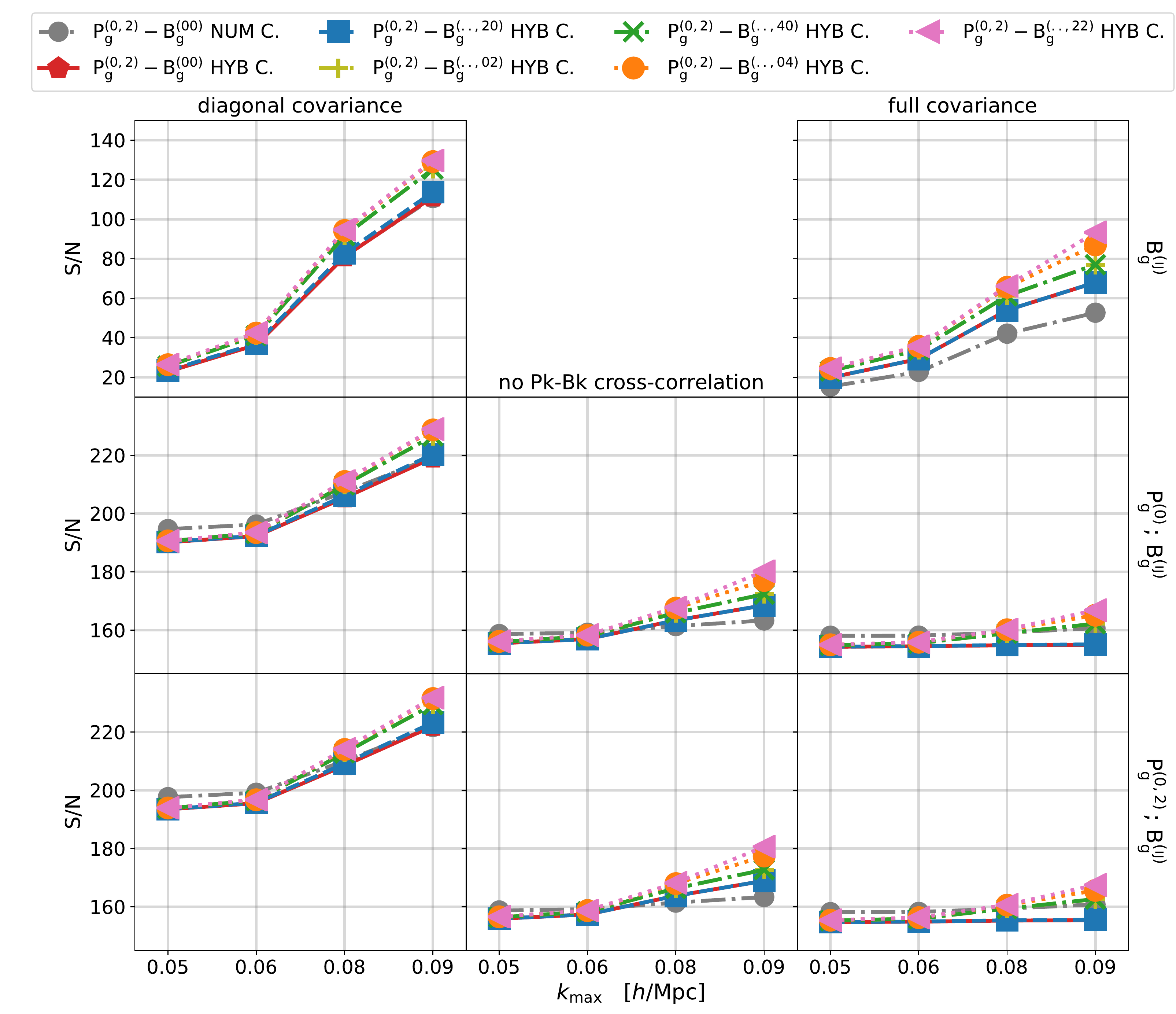}
\caption{\label{fig:ana_dv_sn_ratio_d6}
Comulative signal to noise ratio as a function of $k_\mathrm{max}$ for the different statistics computed using an analytical model for the data vector and both the analytical and numerical covariance matrices, computed using Equation \ref{eq:signal_noise}. In the first row are displayed the lines for the signal to noise ratio when only the bispectrum multipoles are employed. In the second and third rows the power spectrum monopole and quadrupole are added to the data vector, respectively. In the first column only the diagonals of the covariance matrices (uncorrelated data vector's elements) have been used in the cumulative signal to noise ratio computation. In the second column we set to zero the cross-correlation between power spectrum and bispectrum multipoles, leaving the off-diagonal terms in the auto-covariances. Finally in the third column we employed the full covariance matrices to compute the signal to noise ratio. 
From the plot it appears that in terms of cumulative signal to noise the saturation is reached when both the hexadecapole terms of the galaxy bispectrum are included in the data vector.
The discrepancy between the  $\mathrm{\left[\mathrm{P}_\mathrm{g}^{(0,2)};\mathrm{B}_\mathrm{g}^{(00)}\right]}$ lines for numerical and analytical covariance matrices is clearly due to the cross-covariance term between power spectrum and bispectrum multipoles. A possible explanation is the lack of higher order terms in the cross-covariance and the presence of many elements by definition equal to zero whilst not null in the numerical covariance matrix, as visible in the right panel of Figure \ref{fig:covariance_comparison1D}.}
\end{figure}

\subsection{Cumulative signal to noise ratio}
A first preliminary study can be done by looking at the comulative signal to noise ratio (S/N) as a function of the range of scale used (varying $k_\mathrm{max}$ in our case). Given a data vector $\mathbf{x}$ and its covariance matrix $\mathrm{C}_\mathbf{x}$, the squared cumulative signal to noise ratio is given by

\begin{eqnarray}
\label{eq:signal_noise}
\left(\mathrm{S/N}\right)^2\quad = \quad \mathbf{x}^\intercal\;\mathrm{C}_\mathbf{x} \; \mathbf{x}\,.
\end{eqnarray}

\noindent By removing from the bispectrum data vector those triangles having at least one side larger that the chosen $k_\mathrm{max}$ we show in Figure \ref{fig:ana_dv_sn_ratio_d6} the cumulative signal to noise ratio as a function of the considered scales range.

 In Figure \ref{fig:ana_dv_sn_ratio_d6} we look into the signal to noise ratio for different combinations of the data vector's terms (by rows) and different versions of the covariance matrices (by columns). The first message that this plot conveys is that the comulative signal to noise ratio seems to saturate when all three bispectrum hexadecapoles terms are employed in the analysis. Moreover, moving from left to right from the first to second column, it results evident that the overall decrease in signal to noise is caused by the addition of the off-diagonal terms to the covariance matrix. In comparison, adding the cross-correlation between power spectrum and bispectrum has a smaller impact (from second to third column).

 On the third row on the right, when the whole covariance is used (off-diagonal terms and cross-covariances between power spectrum and bispectrum multipoles) there is a  small discrepancy between the (S/N) ratios computed using the numerical and the analytical covariances. The reason for this could be the fact that, as we previously saw in Figure \ref{fig:covariance_comparison1D}, in both the analytically computed bispectrum covariance and cross-covariance terms certain elements are by definition (Kronecker's deltas relations) set to zero, while, because of noise, they are not zero in the numerical covariance. 

\begin{figure}[tbp]
\centering 
\includegraphics[width=.98\textwidth]
{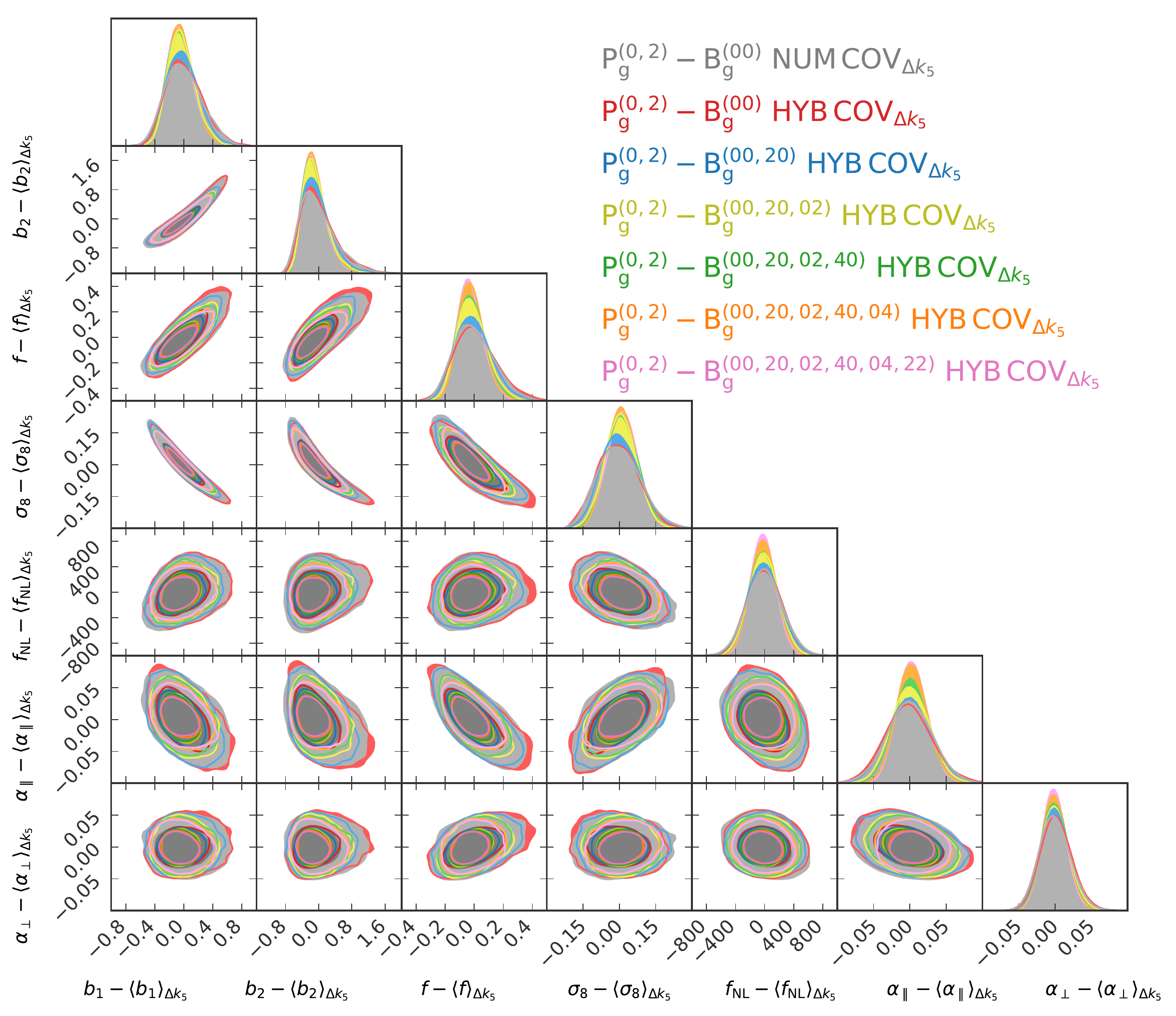}
\caption{\label{fig:ana_dv_6p_d5_improv02}
1- and 2-D marginalised  parameters posterior distributions for different combinations of the bispectrum multipoles, $\Delta k_5$ case. Here we  use the analytical model for a fiducial cosmology as data vector measurement. The gray (NUM COV) and red (HYB COV) contours are obtained using the same statistics (up to the bispectrum monopole) and covariance matrices as in Figure \ref{fig:p02b00p6dk5}. The blue contours are obtained by adding to the data vector the quadrupole term relative to the angle between to the normal to the surface of the triangle and the line of sight ( $\mathrm{\left[\mathrm{P}_\mathrm{g}^{(0,2)};\mathrm{B}_\mathrm{g}^{(00,20)}\right]}$). The yellow contours are obtained by including also the quadrupole term relative to the angle describing the rotation of the triangle around the normal to its surface ( $\mathrm{\left[\mathrm{P}_\mathrm{g}^{(0,2)};\mathrm{B}_\mathrm{g}^{(00,20,02)}\right]}$). In the same way the green and orange contours show the effect of adding the hexadecapoles terms of the two angles in the same order ($\mathrm{\left[\mathrm{P}_\mathrm{g}^{(0,2)};\mathrm{B}_\mathrm{g}^{(00,20,02,40)}\right]}$ and $\mathrm{\left[\mathrm{P}_\mathrm{g}^{(0,2)};\mathrm{B}_\mathrm{g}^{(00,20,02,40,04)}\right]}$), respectively. The addition of the last hexadecapole term (quadrupoles of the two angles used for the polynomial expansion) is shown in pink contours ($\mathrm{\left[\mathrm{P}_\mathrm{g}^{(0,2)};\mathrm{B}_\mathrm{g}^{(00,20,02,40,04,22)}\right]}$). The main message that this figure conveys, together with Table \ref{tab:ana_dv_6p_d6_improv}, is that  significant parameter constraints improvements are obtained up to the bispectrum hexadecapole level.}
\end{figure}

Comparing this to the cumulative signal to noise ratio values displayed in Figure 7 of \cite{Sugiyama:2019ike} we find larger values for the same $k$-range because we used an analytical model as data vector $\mathbf{x}$ while in \cite{Sugiyama:2019ike} the authors used the measurement from the mean of the mocks which contains noise. An example of this can be seen by comparing the contours for the same data vector   $\mathrm{\left[\mathrm{P}_\mathrm{g}^{(0,2)};\mathrm{B}_\mathrm{g}^{(00)}\right]}$ (gray and red) in Figures \ref{fig:p02b00p6dk5} and \ref{fig:ana_dv_6p_d5_improv02} where in the first case the mean of the mocks has been used as data vector measurement while in the second it has been used the analytical model. The 1- and 2-D marginalised contours are much wider in the first case, exactly because of the noise present in a measurement from simulations/data.

\subsection{Improvement on  parameter constraints}
\label{sec:improvement_parameters}

Because of the lack of measurements for the multipoles $ B_g^{(20)}$, $ B_g^{(02)}$, $ B_g^{(40)}$ , $ B_g^{(04)}$ and $ B_g^{(22)}$  we use an analytical  data vector  computed for our fiducial cosmology. 
Figure \ref{fig:ana_dv_6p_d5_improv02} shows the 1 and 2D posterior distributions of the cosmological parameters for data vectors of different lengths, starting from the basic power spectrum monopole and quadruple and bispectrum monopole, higher bispectrum multiplies are gradually added.  Since in the previous tests  comparing analytic/hybrid and numerical  covariance matrix the data vector was taken from the mocks, in Figure \ref{fig:ana_dv_6p_d5_improv02} (gray and red) we  repeat the comparison of sec. \ref{sec:5.2} using analytic the data vector. 
Also in the case of the analytical template for the data vector, the results obtained with numerical and hybrid covariance matrices are in very good agreement. 

The 1-2D marginalised contours shown in Figure \ref{fig:ana_dv_6p_d5_improv02} confirm what deduced from the signal to noise study (Figure \ref{fig:ana_dv_sn_ratio_d6}): saturation is achieved  at the bispectrum hexadecapole level.

We observe a  shrinkage of the 1- and 2-D marginalised posterior distributions as galaxy bispectrum higher multipoles are employed with respect to the monopole term alone. This aspect, together with the saturation, is quantitatively described in Table \ref{tab:ana_dv_6p_d6_improv}.
This analysis suggests that including all the terms up to the galaxy bispectrum hexadecapoles (together with the power spectrum monopole and quadrupole), results in  improving the 1D $68\%$ credible regions by a factor of $\sim 30\%$ on average -- equivalent to observing a volume in the sky $\sim 2 $ times larger ($V_{(\imath\jmath)} \propto 1/\sigma^2_{\imath\jmath} \simeq 1/(0.7\,\sigma_{(00)})^2 = 2 \times V_{(00)}$).

Recall that our analysis is very conservative in terms of scales, by setting $k_\mathrm{max} = 0.09\,h/\mathrm{Mpc}$ we restrict our investigation to the linear regime. Extension to the mildly non-linear regime may offer additional gains. The  cumulative signal to noise shown in Figure \ref{fig:ana_dv_sn_ratio_d6} seems to hint that, as $k_\mathrm{max}$ increases, the difference between the isotropic (monopole) and isotropic plus anistropic (monopole plus higher multipoles) bispectrum signal may also increase.

This trend will eventually stop, as Figure 7 of \cite{Sugiyama:2019ike} seems to indicate around $k_\mathrm{max}\simeq 0.20 \, h/\mathrm{Mpc}$, however it would leave a non negligible gap between the isotropic and anisotropic signal to noise ratios. Therefore as smaller and smaller scales are considered, the signal to noise difference would likely translate into a stronger impact of the anisotropic bispectrum signal in terms of parameter constraints improvements. Full investigation in this direction  is left for future work.

\renewcommand{\arraystretch}{2}
\begin{table}[tbp]
\centering
\begin{tabular}{|c|cccccccc|}
\hline
statistics &  \multicolumn{8}{c|}{$\dfrac{ \Delta\theta^{\mathrm{num.\,cov.}}_{\mathrm{P_g^{(0,2)};B_g^{(00)}}}-\Delta\theta^{\mathrm{hyb.\,cov.}}_{\mathrm{P_g^{(0,2)};B_g^{(\imath,\jmath)}}}}{\Delta\theta^{\mathrm{num.\,cov.}}_{\mathrm{P_g^{(0,2)};B_g^{(00)}}}}\;\left[\%\right]$} \\ [0.5cm]
\hline
$\mathrm{P}_\mathrm{g}^{(0,2)}\;+$  & $b_1$ & $b_2$ & $f$  
& $\sigma_8$ & $f_\mathrm{NL}$  & $\alpha_\parallel$ & $\alpha_\perp$ & average\\ 
\hline   
$\mathrm{\mathrm{B}_\mathrm{g}^{(00)}}$
& 2.6	  & 3.0	 & -5.0	 & 3.1	 & 0.4	 & -6.6 & -0.7 & -0.5	 \\ 
$\mathrm{\mathrm{B}_\mathrm{g}^{(00,20)}}$
& 13.0 & 17.2& 10.7& 11.8&  7.2&  2.7&  5.0 &  9.7\\ 
$\mathrm{\mathrm{B}_\mathrm{g}^{(00,20,02)}}$
& 27.8& 31.2& 25.7& 27.1& 16.8& 14.8&  8.5& 21.7	\\ 
$\mathrm{\mathrm{B}_\mathrm{g}^{(00,20,02,40)}}$
& 26.9& 30.8& 28.0 & 25.2& 17.0 & 19.3& 11.4& 22.7\\ 
$\mathrm{\mathrm{B}_\mathrm{g}^{(00,20,02,40,04)}}$
& \textbf{31.0} & \textbf{34.1} & 36.5& \textbf{30.1} & 25.3 & 31.0 & 18.2& 29.5 \\ 
$\mathrm{\mathrm{B}_\mathrm{g}^{(00,20,02,40,04,22)}}$
& 29.9& 33.6& \textbf{36.6} & 28.4& \textbf{28.5} & \textbf{32.1} & \textbf{21.2} & \textbf{30.0}  \\ 
\hline
\end{tabular}
\caption{\label{tab:ana_dv_6p_d6_improv} Relative improvements on the 1D $68\%$ confidence intervals for the model parameters given by employing the quadrupole and hexadecapole terms of the galaxy bispectrum. In accordance with what observed for the cumulative signal to noise ratio and the 1-2D marginalised posterior distributions in Figures \ref{fig:ana_dv_sn_ratio_d6} and \ref{fig:ana_dv_6p_d5_improv02} respectively, we see that the saturation in terms of information is reached with the addition of the third bispectrum hexadecapole term (the largest constraints improvements have been highlighted using the bold font). As an additional check on the analytical data vector, the first row shows that there is a negligible difference between the constraints obtained using the analytical and numerical covariance matrix for the same data vector.}
\end{table}

\begin{figure}[tbp]
\centering 
\includegraphics[width=.95\textwidth]
{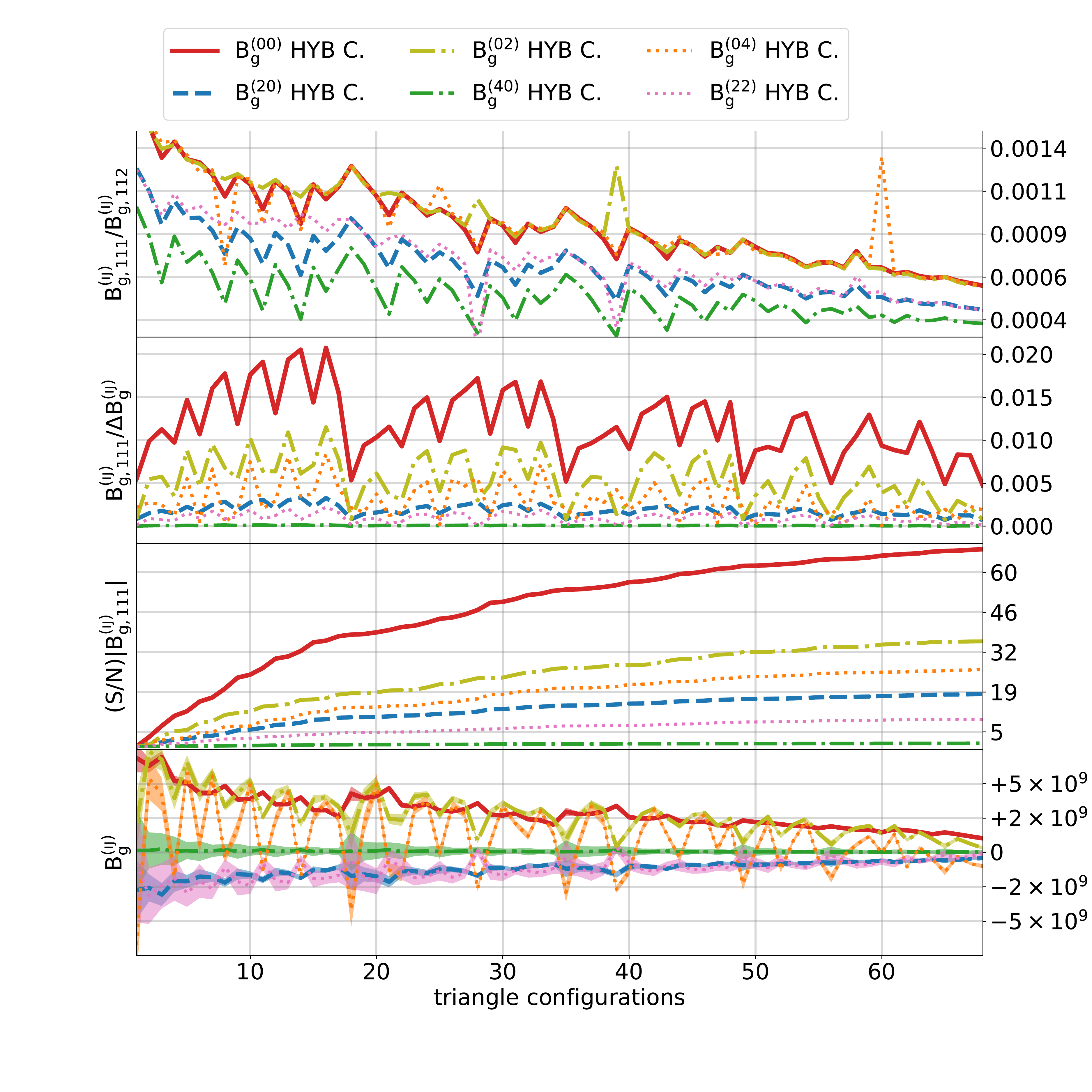}
\caption{\label{fig:png_higher_multi}Primordial non-Gaussianity signal in the bispectrum multipoles for $f_\mathrm{NL}=5$. As indicated in the legend each line color/linestyle, corresponds to a different bispectrum multipole. The top panel shows the ratio between the primordial ($B^{(ij)}_{\mathrm{g},1,1,1}$) and the gravitational parts ($B^{(ij)}_{\mathrm{g},1,1,2}$).
 In the second panel from the top we show  the ratio between each multipole's primordial component and the variance given by the calibrated hybrid covariance matrix is displayed. This can be interpreted as a signal-to-noise ratio. In the third row, the cumulative signal to noise ratio of the primordial non-Gaussian bispectrum component is reported.
Finally in the bottom panel the full bispectrum signal (primordial plus gravitational) for each multipole is plotted, always as a function of the triangle configurations. Once more the shaded regions correspond to the square root of the analytical covariance matrices diagonal elements.}
\end{figure}

\subsection{Local Primordial non-Gaussianity}

The bispectrum has been widely studied in the literature \cite{Verde:1999ij,Komatsu:2001rj,Sefusatti:2009qh,Jeong:2009vd,Scoccimarro:2003wn} as a powerful tool to constrain potential deviations from Gaussianity of the primordial gravitational potential \cite{Bardeen:1980kt}.
By reducing the degeneracies between cosmological parameters, the galaxy bispectrum higher multipoles can play an important role in obtaining late-times constraints on the primordial non-Gaussianity amplitude parameter $f_\mathrm{NL}$ \cite{Bartolo:2004if} competitive with the early times ones produced by cosmic microwave background (CMB) experiment such as Planck \cite{Akrami:2019izv}.
Moreover constraining $f_\mathrm{NL}$ using the bispectrum multipoles does not rely on extremely large scales and on highly biased tracers as the current state-of-the-art analysis of the power spectrum scale dependent bias \cite{Castorina:2019wmr}.

In our estimate of the signature of primordial non-Gaussianity in the anisotropic bispectrum we do not include the modelling of the terms associated with the scale dependent bias induced by a local $f_\mathrm{NL}$ \cite{Dalal:2007cu,Matarrese:2008nc} and arising from the multivariate bias expansion \cite{Giannantonio:2009ak,Baldauf:2010vn,Tellarini:2015faa} applied to the bispectrum. This effect is particularly strong at very large scales and an analysis aimed at deriving constraints on the local non-Gaussianity amplitude parameter $f_\mathrm{NL}$ using the bispectrum should implement the modelling developed in \cite{Tellarini:2016sgp,Karagiannis:2018jdt} which also includes the multivariate biasing expansion in redshift-space.
Our analysis can be interpreted as a conservative estimate using only the primordial signal appearing from a non vanishing initial bispectrum, underestimating the actual signature given by $f_\mathrm{NL}\neq0$. This would be larger and likely less correlated with the other cosmological parameters, especially on large scales, if we would include the non-local bias effect.
This is possibly one of the reasons why we obtain larger confidence intervals for $f_\mathrm{NL}$ than the forecasts made through a Fisher information analysis in \cite{Tellarini:2016sgp}.

In particular redshift space distortions add an additional difference in the standard perturbation theory kernels dependence between the gravitational collapse ($\mathrm{B}^{(\imath\jmath)}_{\mathrm{g},112}\propto \mathrm{Z^\mathrm{s}_1}\mathrm{Z^\mathrm{s}_1}\mathrm{Z^\mathrm{s}_2}$)  and primordial ($\mathrm{B}^{(\imath\jmath)}_{\mathrm{g},111}\propto \mathrm{Z^\mathrm{s}_1}\mathrm{Z^\mathrm{s}_1}\mathrm{Z^\mathrm{s}_1}$) components of the bispectrum which becomes more explicit when the anisotropic signal is studied.

In Figure \ref{fig:png_higher_multi} we show the local primordial non-Gaussianity contributions to the bispectrum \cite{Byrnes:2010em} for the different multipoles, setting $f_\mathrm{NL}=5$. These terms have been computed using the expression reported in the Appendix of \cite{Gualdi:2019sfc}. The first row shows, for the different multipoles, the ratio between the primordial signal\footnote{It is interesting to notice that the multipoles that appear to be more sensitive to (richer in signal for) local primordial non-Gaussianity, are the ones relative to the angle describing the triangle rotation around the normal to the surface. An hint for a potential explanation to this effect could be searched in the angle conversion Equation \ref{eq:angles_conversion}. The cosine $\nu$ of the angle describing the rotation of the triangle around the normal to the surface depends explicitly on $\cos\alpha_{ab}$. This is the cosine of the internal angle of the triangle projection on the plane perpendicular to the line of sight. Therefore $\nu$ explicitly depends on how much squeezed is the triangle, which is a factor known to influence the sensitivity to primordial non-Gaussianity \cite{Figueroa:2012ws,Baldauf:2010vn,Byun:2013jba,Liguori:2010hx}.} 
$\mathrm{B^{(\imath\jmath)}_{g,111}}$ and the gravitational collapse part of the bispectrum $\mathrm{B^{(\imath\jmath)}_{g,112}}$.
In the second panel from the top, the ratio is between $\mathrm{B^{(\imath\jmath)}_{g,111}}$ and the variance computed using the analytical covariance matrix model used in the paper's analysis. On the third row, the comulative signal-to-noise ratio is shown as a function of the number of triangle configurations included in the data vector. Finally in the bottom panel it is possible to see the  total signal (primordial plus gravitational) for the different multipoles as a function of the triangle configurations considered.

\section{Conclusions}
\label{sec:conclusions}

At the power spectrum level, it is well known that  the anisotropic modulation of the signal with respect to the line-of-sight induced by redshift space distortions  carries valuable cosmological information. This anisotropy is usually expressed in multipole moments: the monopole  --being the averaged signal in all directions around the line of sight-- and the quadruple. 
The bispectrum has been proved useful for breaking degeneracies among parameters and detecting specific physical signatures (e.g., the so-called GR effects or neutrino masses). To date, only the bispectrum monopole has been used to constrain cosmological parameters from galaxy spectroscopic surveys. Measuring higher-order multipoles of the redshift-space galaxy bispectrum is indeed challenging and computationally expensive \cite{Scoccimarro:2015bla,Hashimoto:2017klo,Sugiyama:2018yzo}. In this paper we have addressed the question: given these challenges, is it worth it?
We consider that  the answer is "yes", if adding the bispectrum anisotropic signal to the standard power spectrum $\mathrm{P}_\mathrm{g}^{(\ell)}$ (isotropic + anisotropic) and bispectrum $\mathrm{B}_\mathrm{g}^{(00)}$ (isotropic) data vector, produces tighter 1- and 2-D marginalised posterior distributions for the parameters usually constrained by galaxy-clustering analyses.

To do so, we developed  an analytical model for the  full covariance matrix including power spectrum monopole and quadruple and bispectrum monopole and higher-orders multipoles up to the hexadecapole, for a conservative range of scales ($k_\mathrm{max}=0.09\,h/\mathrm{Mpc}$). We concentrated on the set of cosmological parameters  $(b_1,b_2,f,\sigma_8,f_\mathrm{NL},\alpha_\perp, \alpha_\parallel)$.

We first calibrate the analytical covariance matrix on the one estimated  numerically from a suite of 1400 mock survey catalogs. The calibration is done on the sub-matrix corresponding to the reduced data vector $\mathrm{\left[\mathrm{P}_\mathrm{g}^{(0,2)};\mathrm{B}_\mathrm{g}^{(00)}\right]}$. We have shown that  the parameter's 1- and 2-D marginalised  posteriors distributions obtained with our analytical covariance and the numerical one are very similar and basically indistinguishable for  practical applications of reporting error-bars on parameters values (Figures  \ref{fig:p02b00p6dk5}, \ref{fig:stat_test_p02b00_violin}, \ref{fig:stat_test_p02b00_bfdistribution}).

Forecasts of the improvements in terms of parameter constraints induced by adding to the data vector the galaxy bispectrum anisotropic signal were presented in Section \ref{sec:higher_bk_multi}. These have been obtained by running standard MCMC samplings together with using the analytical covariance matrix to evaluate the likelihood.
For a BOSS-like survey, we found on average $\sim 22 \%$ tighter 1D $68\%$ credible regions by just considering the two bispectrum quadrupoles in addition to the monopole (Figure \ref{fig:ana_dv_6p_d5_improv02} and Table \ref{tab:ana_dv_6p_d6_improv}).

Including the last galaxy bispectrum hexadecapole term $\mathrm{B_g}^{(22)}$ only returns a further smaller improvement, suggesting that the constraining power of the bispectrum saturates at this level when expanded in terms of Legendre polynomials. This hypothesis is supported by the behaviour  of the  signal to noise ratio in Figure \ref{fig:ana_dv_sn_ratio_d6} and 1-2D marginalised posterior distributions (again Figure \ref{fig:ana_dv_6p_d5_improv02} and Table \ref{tab:ana_dv_6p_d6_improv}), where adding the third hexadecapole term, $\mathrm{B_g^{(22)}}$, does not provide any significant information gain.
Improving by $\sim 30\%$ on average the parameter constraints by employing the bispectrum anisotropic signal approximately corresponds to observing a  survey volume two times larger for a BOSS-like survey.

We stress that these results are conservative since we restrict our analysis to linear scales $k_\mathrm{max} =0.09\,h$/Mpc. Cumulative signal to noise studies in the literature indicate that the benefits of both employing the anisotropic bispectrum \cite{Sugiyama:2019ike} and adding the bispectrum to the power spectrum analyses \cite{Sefusatti:2004xz,Byun:2017fkz,Yankelevich:2018uaz}, increase for larger $k_\mathrm{max}$.

In Appendix \ref{sec:deltak6} we report the key Figures displaying the results obtained when the same analysis done in the main body of the paper is also performed on the bispectrum data vector obtained for a larger $k$-bin width ($\Delta k_6$), hence with less triangle configurations.
Even if the overall improvement on the parameters constraints appears to be larger, information gain saturation is again reached at the bispectrum hexadecapole level. Moreover, comparing the 1- and 2-D marginalised posterior distributions for the two $k$-bin width cases, Figure \ref{fig:ana_dv_6p_d5_improv02} for $\Delta k_5$ and Figure \ref{fig:ana_dv_6p_d6_improv02} for $\Delta k_6$, we appreciate that  the contours for the full data vector including all the multipoles up to the hexadecapole level are qualitatively very similar.

We  therefore conclude that the (significant) effort necessary to include the bispectrum quadruple in clustering analysis of galaxy redshift surveys is very likely to pay off.  We envision that this extra information could be used not only to reduce parameter errors; these additional statistics can provide a useful consistency checks of the modelling itself but possibly also of the model.

\appendix

\section{Covariance matrix terms derivation}
In this Appendix are reported the full derivations of the covariance terms used in this work. We build on and improve what already presented in a previous paper \cite{Gualdi:2018pyw}. In particular we are interested in computing the covariance expressions for both power spectrum and bispectrum expansion in terms of Legendre polynomials encoding the relation between the signal and the $k$-vectors orientation with respect to the line of sight.

\label{sec:app_cov_terms}
\subsection{Power spectrum multipoles auto-covariance}
In order to compute the covariance of the power spectrum multipoles it is necessary to evaluate the four-point correlator in Fourier space:

\begin{align}
\label{eq:pk_cov_expansion}
\langle\delta_{\mathrm{g}}(\mathbf{k}_1)\delta_{\mathrm{g}}(-\mathbf{k}_1)\delta_{\mathrm{g}}(\mathbf{k}_2)\delta_{\mathrm{g}}(-\mathbf{k}_2) \rangle
&= \langle\delta_{\mathrm{g}}(\mathbf{k}_1)\delta_{\mathrm{g}}(-\mathbf{k}_2)\rangle_{\mathrm{c}} \langle\delta_{\mathrm{g}}(\mathbf{k}_2)\delta_{\mathrm{g}}(-\mathbf{k}_1)\rangle_{\mathrm{c}} +\,1\,\mathrm{perm.}
\notag\\
&+\langle\delta_{\mathrm{g}}(\mathbf{k}_1)\delta_{\mathrm{g}}(\mathbf{k}_2)\delta_{\mathrm{g}}(-\mathbf{k}_1)\delta_{\mathrm{g}}(-\mathbf{k}_2)\rangle_{\mathrm{c}}
,
\end{align}
\noindent where the term in the  second line is proportional to the trispectrum for parallelogram configurations. Since we limit our analysis to linear scales, we only use the first Gaussian term, which dominates over the non-Gaussian one. For a recent work regarding the power spectrum covariance matrix see \cite{Wadekar:2019rdu}. 
It is possible to compute the Gaussian term of the covariance for the power spectrum monopole and quadrupole as follows:

\begin{align}
&\mathrm{C}_\mathrm{G}^{\mathrm{P}^{(\ell)}_{\mathrm{g}}\mathrm{P}^{(\ell)}_{\mathrm{g}}}\left(k_1;k_2\right) = 
\left(2\ell+1\right)^2\dfrac{(2\pi)^{-6}}{N_{\mathrm{p}}\left(k_1\right)N_{\mathrm{p}}\left(k_2\right)}
\int_{V_{\mathbf{q}_1}}\int_{V_{\mathbf{q}_2}}\int_{V_{\mathbf{p}_1}}\int_{V_{\mathbf{p}_2}}
 d^3\mathbf{q}_1 d^3\mathbf{q}_2 d^3\mathbf{p}_1 d^3\mathbf{p}_2
 \notag \\
&\times
\mathcal{L}_{\ell}\left(\mu_1\right)\mathcal{L}_{\ell}\left(\mu_2\right)
\delta_D\left(\mathbf{q}_1+\mathbf{p}_1\right)\delta_D\left(\mathbf{q}_2+\mathbf{p}_2\right) 
2(2\pi)^6\delta_D\left(\mathbf{q}_1+\mathbf{q}_2\right)\delta_D\left(\mathbf{p}_1+\mathbf{p}_2\right)
\mathrm{P}_{\mathrm{g}}^{\mathrm{s}}\left(\mathbf{q}_1\right)\mathrm{P}_{\mathrm{g}}^{\mathrm{s}}\left(\mathbf{p}_2\right)
\notag \\
&=
\left(2\ell+1\right)^2\dfrac{2}{N_{\mathrm{p}}\left(k_1\right)N_{\mathrm{p}}\left(k_2\right)}
\int_{V_{\mathbf{q}_1}}\int_{V_{\mathbf{q}_2}}
 d^3\mathbf{q}_1 d^3\mathbf{q}_2
\mathcal{L}_{\ell}\left(\mu_1\right)\mathcal{L}_{\ell}\left(\mu_2\right)
\delta_D\left(\mathbf{q}_1+\mathbf{q}_2\right)^2
\mathrm{P}_{\mathrm{g}}^{\mathrm{s}}\left(\mathbf{q}_1\right)\mathrm{P}_{\mathrm{g}}^{\mathrm{s}}\left(\mathbf{q}_2\right)
\notag \\
&=
\left(2\ell+1\right)^2\dfrac{2}{N_{\mathrm{p}}\left(k_1\right)V_{\mathrm{p}}\left(k_2\right)}
\int_{V_{\mathbf{q}_1}}\int_{V_{\mathbf{q}_2}}
 d^3\mathbf{q}_1 d^3\mathbf{q}_2
\mathcal{L}_{\ell}\left(\mu_1\right)\mathcal{L}_{\ell}\left(\mu_2\right)
\delta_D\left(\mathbf{q}_1+\mathbf{q}_2\right)
\mathrm{P}_{\mathrm{g}}^{\mathrm{s}}\left(\mathbf{q}_1\right)\mathrm{P}_{\mathrm{g}}^{\mathrm{s}}\left(\mathbf{q}_2\right)
\notag \\
&\approx 
\left(2\ell+1\right)^2\dfrac{2k_f^{3}}{V_{\mathrm{p}}\left(k_1\right)}\quad
\dfrac{1}{2}\int^{1}_{-1}d\mu_1 \mathcal{L}_\ell\left(\mu_1\right)^2
\mathrm{P}_{\mathrm{g}}\left(k_1,\mu_1\right)^2
\notag \\
&=
\left(2\ell+1\right)^2
\dfrac{2k_f^{3}\delta^\mathrm{K}_{12}}{4\pi k_1^2\Delta k}\quad
\dfrac{1}{2}\int^{1}_{-1}d\mu_1 \mathcal{L}_\ell\left(\mu_1\right)^2
\mathrm{P}_{\mathrm{g}}\left(k_1,\mu_1\right)^2,
\end{align}

\noindent where again we used the approximation made in \cite{Joachimi:2009zj} that $\delta_{\mathrm{D}}^2\approx\dfrac{V_{\mathrm{s}}}{(2\pi)^3}\delta_{\mathrm{D}} = k_f^{-3}\delta_{\mathrm{D}}$. $V_\mathrm{s}$ is the survey volume $k_f$ the fundamental frequency.  $\delta^{\mathrm{K}}_{12}$ is the Kronecker delta indicating that the vector $\mathbf{q}_1$ and $\mathbf{q}_2$ are identical (in the second step trivial $\delta_{\mathrm{K}}$ have been omitted in order to avoid making the notation heavier by adding also the wave-vector letter).  The Fourier integration volume for the power spectrum is defined as $V_\mathrm{p} = 4\pi\Delta k k^2 $ and therefore the number of modes ("pairs") as $N_\mathrm{p} = V_\mathrm{p} / k_f^3 $. ${\cal L}$ denotes the Lagrange polynomials. In the last steps we made the approximation that the power spectrum monopole and quadrupoles do not significantly vary when integrated over the bin in Fourier space.

\subsection{Bispectrum multipoles auto-covariance}
\label{sec:appendixB2}
We write these expression as functions of the multipoles of two angles defined in \cite{Hashimoto:2017klo}. The first is the angle $\omega$ between the normal to the surface of the triangle and the line of sight. The second is the angle $\phi$ between the first k-vector and one of the two axis defining the plane on which the triangle lies. We will then use the cosines for the multipoles $\mu = \cos\omega$ and $\nu = \cos\phi$. The multipoles indexes will be given then by $\mathcal{L}^{(\alpha)}_{\mu_a}$, $\mathcal{L}^{(\beta)}_{\nu_a}$ for the first triangle and $\mathcal{L}^{(\kappa)}_{\mu_b}$, $\mathcal{L}^{(\lambda)}_{\nu_b}$ for the second one. We introduce the short notation $\mathcal{M}^{(\alpha\beta\kappa\lambda)}_{\mu\nu,ab} =\mathcal{L}^{(\alpha)}_{\mu_a}\mathcal{L}^{(\beta)}_{\nu_a}\mathcal{L}^{(\kappa)}_{\mu_b}\mathcal{L}^{(\lambda)}_{\nu_b} $. In the same way we  will write the compact pre-factor $\mathcal{C}^{\alpha\beta}_{\kappa\lambda} = (2\alpha+1)(2\beta+1)(2\kappa+1)(2\lambda+1)$.

Following the angles definition in \cite{Hashimoto:2017klo}, the conversion from the standard bispectrum coordinates $(k_1,k_2,k_3,\mu_1,\mu_2)$ to the set used in this work  $(k_1,k_2,k_3,\mu,\nu)$ is given by

\begin{align}
\label{eq:angles_conversion}
        \mu = \cos\omega &= \sqrt{\dfrac{2\left(a_\perp^2b_\perp^2 + a_\perp^2c_\perp^2 + b_\perp^2c_\perp^2\right) - a_\perp^4 - b_\perp^4 - c_\perp^4}{2\left(a^2b^2 + a^2c^2 + b^2c^2\right)-a^4-b^4-c^4}} = \sqrt{\dfrac{\Delta_\perp}{\Delta}}
    \notag \\
    \nu = \cos\phi & = \dfrac{\left(1-\mu_a^2\right)\mu_b + \cos\alpha_{ab}\mu_a\sqrt{1-\mu_a^2}\sqrt{1-\mu_b^2}}{\sqrt{1-\mu^2}} \,,
\end{align}{}

\noindent where for notation simplicity we used $(a,b,c)$ instead of $(k_1,k_2,k_3)$. The '$\perp$' indicates the 2D vector obtained by projecting the 3D $k$-vector onto to the plane perpendicular to the line of sight. $\alpha_{ab}$ is the internal angle between the sides $a_\perp$ and $b_\perp$ of the $a_\perp b_\perp c_\perp$ triangle. When $|\mu| = 1$ the triangle normal to the surface is either parallel or anti-parallel to the line of sight. Therefore the anisotropic bispectrum signal is the same for all the potential values of $\nu$ since the triangle plane coincides with the one perpendicular to the line of sight.

In order to compute the covariance between two galaxy bispectrum multipoles, the six-points correlator in Fourier space given by the respective galaxy density variables needs to be evaluated \cite{Sefusatti:2006pa}:

\begin{align}
\label{eq:bk_cov_expansion}
\langle&\delta_{\mathrm{g}}(\mathbf{k}_1)\delta_{\mathrm{g}}(\mathbf{k}_2)\delta_{\mathrm{g}}(\mathbf{k}_3)\delta_{\mathrm{g}}(\mathbf{k}_4)\delta_{\mathrm{g}}(\mathbf{k}_5)\delta_{\mathrm{g}}(\mathbf{k}_6) \rangle
\notag \\
&= \langle\delta_{\mathrm{g}}(\mathbf{k}_1)\delta_{\mathrm{g}}(\mathbf{k}_4)\rangle_{\mathrm{c}} \langle\delta_{\mathrm{g}}(\mathbf{k}_2)\delta_{\mathrm{g}}(\mathbf{k}_5)\rangle_{\mathrm{c}}
\langle\delta_{\mathrm{g}}(\mathbf{k}_3)\delta_{\mathrm{g}}(\mathbf{k}_6)\rangle_{\mathrm{c}} +\,8\,\mathrm{perms.}
\notag\\
&+\langle\delta_{\mathrm{g}}(\mathbf{k}_1)\delta_{\mathrm{g}}(\mathbf{k}_2)\delta_{\mathrm{g}}(\mathbf{k}_4)\rangle_{\mathrm{c}}
\langle\delta_{\mathrm{g}}(\mathbf{k}_3)\delta_{\mathrm{g}}(\mathbf{k}_5)\delta_{\mathrm{g}}(\mathbf{k}_6)\rangle_{\mathrm{c}} +\,8\,\mathrm{perms.}
\notag\\
&+\langle\delta_{\mathrm{g}}(\mathbf{k}_1)\delta_{\mathrm{g}}(\mathbf{k}_2)\delta_{\mathrm{g}}(\mathbf{k}_5)\delta_{\mathrm{g}}(\mathbf{k}_6)\rangle_{\mathrm{c}}\langle\delta_{\mathrm{g}}(\mathbf{k}_3)\delta_{\mathrm{g}}(\mathbf{k}_4)\rangle_{\mathrm{c}}+\,8\,\mathrm{perms.}
\notag\\
&+\langle\delta_{\mathrm{g}}(\mathbf{k}_1)\delta_{\mathrm{g}}(\mathbf{k}_2)\delta_{\mathrm{g}}(\mathbf{k}_3)\delta_{\mathrm{g}}(\mathbf{k}_4)\delta_{\mathrm{g}}(\mathbf{k}_5)\delta_{\mathrm{g}}(\mathbf{k}_6)\rangle_{\mathrm{c}}.
\end{align}

\noindent From the above expression the bispectrum covariance results composed by four terms. The first is the so called Gaussian term which being proportional to the cube of the power spectrum represents the lowest order in the perturbation expansion ($\propto\delta_\mathrm{m}^6$ where "m" indicates the matter density field). The second and third terms are the lowest order non-Gaussian contributions ($\propto\delta_\mathrm{m}^8$), proportional to the product of two bispectra and power spectrum times trispectrum, respectively.
The last is proportional to the pentaspectrum and since it is an higher order term than the previous ones ($\propto\delta_\mathrm{m}^{10}$), it will not be computed explicitly.

\paragraph{Gaussian term:}
Here we update and extend  the work presented in \cite{Gualdi:2018pyw}.
Analogously to what done  above, we now compute the diagonal term of the bispectrum multipoles covariance matrix:
\begin{align}
\label{eq:cov_bb_g}
&\mathrm{C}_{\propto\mathrm{P}^3}^{\mathrm{B}^{\alpha\beta}_{\mathrm{g}}\mathrm{B}^{\kappa\lambda}_{\mathrm{g}}}\left(k_1,k_2,k_3;k_4,k_5,k_6\right) = 
\notag \\
&=
\dfrac{(2\pi k_f)^{-6}\mathcal{C}^{\alpha\beta}_{\kappa\lambda} }{N^{\mathrm{t}}_{123}N^{\mathrm{t}}_{456}}
\prod^6_{i=1}\int_{V_{\mathbf{q}_i}} d^3\mathbf{q}_i
\mathcal{M}^{(\alpha\beta\kappa\lambda)}_{\mu\nu,ab}
\delta_D\left(\mathbf{q}_1+\mathbf{q}_2+\mathbf{q}_3\right)\delta_D\left(\mathbf{q}_4+\mathbf{q}_5+\mathbf{q}_6\right)
\notag\\
&\times
(2\pi)^9\delta_D\left(\mathbf{q}_1+\mathbf{q}_4\right)\delta_D\left(\mathbf{q}_2+\mathbf{q}_5\right)\delta_D\left(\mathbf{q}_3+\mathbf{q}_6\right)
\mathrm{P}_{\mathrm{g}}^{\mathrm{s}}\left(\mathbf{q}_1\right)\mathrm{P}_{\mathrm{g}}^{\mathrm{s}}\left(\mathbf{q}_2\right)\mathrm{P}_{\mathrm{g}}^{\mathrm{s}}\left(\mathbf{q}_3\right) \; + 5 \;\mathrm{perm.}
\notag \\
&=
\dfrac{\mathrm{D}^{123}_{456}(2\pi)^3 k_f^{-6}\mathcal{C}^{\alpha\beta}_{\kappa\lambda}}
{N^{\mathrm{t},2}_{123}}
\prod^3_{i=1}\int_{V_{\mathbf{q}_i}} d^3\mathbf{q}_i
\mathcal{M}^{(\alpha\beta\kappa\lambda)}_{\mu\nu,aa}
\delta_D\left(\mathbf{q}_1+\mathbf{q}_2+\mathbf{q}_3\right)^2
\mathrm{P}_{\mathrm{g}}^{\mathrm{s}}\left(\mathbf{q}_1\right)\mathrm{P}_{\mathrm{g}}^{\mathrm{s}}\left(\mathbf{q}_2\right)\mathrm{P}_{\mathrm{g}}^{\mathrm{s}}\left(\mathbf{q}_3\right)
\notag \\
&=
\dfrac{\mathrm{D}^{123}_{456}(2\pi)^3 k_f^{-9}\mathcal{C}^{\alpha\beta}_{\kappa\lambda}}
{N^{\mathrm{t},2}_{123}}
\prod^3_{i=1}\int_{V_{\mathbf{q}_i}} d^3\mathbf{q}_i
\mathcal{M}^{(\alpha\beta\kappa\lambda)}_{\mu\nu,aa}
\delta_D\left(\mathbf{q}_1+\mathbf{q}_2+\mathbf{q}_3\right)
\mathrm{P}_{\mathrm{g}}^{\mathrm{s}}\left(\mathbf{q}_1\right)\mathrm{P}_{\mathrm{g}}^{\mathrm{s}}\left(\mathbf{q}_2\right)\mathrm{P}_{\mathrm{g}}^{\mathrm{s}}\left(\mathbf{q}_3\right)
\notag \\
&=
\dfrac{\mathrm{D}^{123}_{456}(2\pi)^3 k_f^{3}\mathcal{C}^{\alpha\beta}_{\kappa\lambda}}
{V^{\mathrm{t},2}_{123}}
\prod^3_{i=1}\int_{V_{\mathbf{q}_i}} d^3\mathbf{q}_i
\mathcal{M}^{(\alpha\beta\kappa\lambda)}_{\mu\nu,aa}
\delta_D\left(\mathbf{q}_1+\mathbf{q}_2+\mathbf{q}_3\right)
\mathrm{P}_{\mathrm{g}}^{\mathrm{s}}\left(\mathbf{q}_1\right)\mathrm{P}_{\mathrm{g}}^{\mathrm{s}}\left(\mathbf{q}_2\right)\mathrm{P}_{\mathrm{g}}^{\mathrm{s}}\left(\mathbf{q}_3\right)
\notag \\
&\approx
\dfrac{\mathrm{D}^{123}_{456} (2\pi)^3k_f^{3}\mathcal{C}^{\alpha\beta}_{\kappa\lambda}}
{V^{\mathrm{t}}_{123}}
\dfrac{1}{4\pi}
\int^{+1}_{-1}d\mu_1\int^{2\pi}_{0}d\phi
\mathcal{M}^{(\alpha\beta\kappa\lambda)}_{\mu\nu,aa}
\mathrm{P}_{\mathrm{g}}\left(k_1,\mu_1\right)\mathrm{P}_{\mathrm{g}}\left(k_2,\mu_2\right)\mathrm{P}_{\mathrm{g}}\left(k_3,\mu_3\right),
\end{align}

\noindent where $\mathrm{D}^{123}_{456}$ stands for all the possible permutations and has values $6,2,1$ respectively for equilateral, isosceles and scalene triangles.
 $V^{\mathrm{t}}_{123} = 8\pi^2k_1k_2k_3\Delta k_1\Delta k_2\Delta k_3$ is the integration volume in Fourier space for the bispectrum. The number of modes is then defined as $N^{\mathrm{t}}_{123}=V^{\mathrm{t}}_{123}/k_f^6$.
Again,  we have  assumed that the power spectrum monopole does not vary significantly inside the integration volume. The angle $\phi$ is defined such as $\mu_2\equiv\mu_1\cos\psi_{12} - \sqrt{1-\mu_1^2}\sqrt{1-\cos\psi_{12}^2}\cos\phi$ where $\cos\psi_{12}\equiv(\mathbf{k}_1\cdot\mathbf{k}_2/(k_1k_2)$. In order to be sure of the symmetry of the numerical result we integrate for each possible combination of $k$-vectors $(k_1,k_2)$, $(k_1,k_3)$ and $(k_2,k_3)$.

%
\paragraph{non-Gaussian terms:}
the following term (Bispectrum cross covariance $\propto BB$) was also presented in \cite{Gualdi:2018pyw} and allows one to better capture the correlation between different triangle bins. It is one of the two non-Gaussian terms that appear in the bispectrum self-covariance matrix up to order $\propto \delta_\mathrm{m}^8$.

\begin{align}
\label{eq:cov_bb_ng}
&\mathrm{C}_{\propto\mathrm{BB}}^{\mathrm{B}^{\alpha\beta}_{\mathrm{g}}\mathrm{B}^{\kappa\lambda}_{\mathrm{g}}}\left(k_1,k_2,k_3;k_4,k_5,k_6\right) = 
\notag\\
&=
\dfrac{(2\pi k_f)^{-6}\mathcal{C}^{\alpha\beta}_{\kappa\lambda} }{N^{\mathrm{t}}_{123}N^{\mathrm{t}}_{456}}
\prod^6_{i=1}\int_{V_{\mathbf{q}_i}} d^3\mathbf{q}_i
\mathcal{M}^{(\alpha\beta\kappa\lambda)}_{\mu\nu,ab}
\delta_D\left(\mathbf{q}_1+\mathbf{q}_2+\mathbf{q}_3\right)\delta_D\left(\mathbf{q}_4+\mathbf{q}_5+\mathbf{q}_6\right)
\notag\\
&\times
(2\pi)^6\delta_D\left(\mathbf{q}_1+\mathbf{q}_2+\mathbf{q}_4\right)\delta_D\left(\mathbf{q}_3+\mathbf{q}_5+\mathbf{q}_6\right)
\mathrm{B}_{\mathrm{g}}^{\mathrm{s}}\left(\mathbf{q}_1,\mathbf{q}_2,\mathbf{q}_4\right)\mathrm{B}_{\mathrm{g}}^{\mathrm{s}}\left(\mathbf{q}_3,\mathbf{q}_5,\mathbf{q}_6\right) \; + 8 \;\mathrm{perm.}
\notag \\
&=
\dfrac{k_f^{-6}\mathcal{C}^{\alpha\beta}_{\kappa\lambda}\delta^{\mathrm{K}}_{34} }{N^{\mathrm{t}}_{123}N^{\mathrm{t}}_{456}}
\int_{V_{\mathbf{q}_1}}\int_{V_{\mathbf{q}_2}}\int_{V_{\mathbf{q}_3}}\int_{V_{\mathbf{q}_5}}\int_{V_{\mathbf{q}_6}}
d^3\mathbf{q}_1d^3\mathbf{q}_2d^3\mathbf{q}_3d^3\mathbf{q}_5d^3\mathbf{q}_6
\,\mathcal{M}^{(\alpha\beta\kappa\lambda)}_{\mu\nu,ab}
\delta_D\left(\mathbf{q}_1+\mathbf{q}_2+\mathbf{q}_3\right)
\notag\\
&\times
\delta_D\left(\mathbf{q}_3+\mathbf{q}_5+\mathbf{q}_6\right)^2
\mathrm{B}_{\mathrm{g}}^{\mathrm{s}}\left(\mathbf{q}_1,\mathbf{q}_2,\mathbf{q}_3\right)\mathrm{B}_{\mathrm{g}}^{\mathrm{s}}\left(\mathbf{q}_3,\mathbf{q}_5,\mathbf{q}_6\right) \; + 8 \;\mathrm{perm.}
\notag \\
&=
\dfrac{k_f^{-9}\mathcal{C}^{\alpha\beta}_{\kappa\lambda}\delta^{\mathrm{K}}_{34} }{N^{\mathrm{t}}_{123}N^{\mathrm{t}}_{456}}
\int_{V_{\mathbf{q}_1}}\int_{V_{\mathbf{q}_2}}\int_{V_{\mathbf{q}_3}}\int_{V_{\mathbf{q}_5}}\int_{V_{\mathbf{q}_6}}
\mathcal{M}^{(\alpha\beta\kappa\lambda)}_{\mu\nu,ab}
d^3\mathbf{q}_1d^3\mathbf{q}_2d^3\mathbf{q}_3d^3\mathbf{q}_5d^3\mathbf{q}_6 \,
\delta_D\left(\mathbf{q}_1+\mathbf{q}_2+\mathbf{q}_3\right)
\notag\\
&\times
\delta_D\left(\mathbf{q}_3+\mathbf{q}_5+\mathbf{q}_6\right)
\mathrm{B}_{\mathrm{g}}^{\mathrm{s}}\left(\mathbf{q}_1,\mathbf{q}_2,\mathbf{q}_3\right)\mathrm{B}_{\mathrm{g}}^{\mathrm{s}}\left(\mathbf{q}_3,\mathbf{q}_5,\mathbf{q}_6\right) \; + 8 \;\mathrm{perm.}
\notag \\
&=
\dfrac{k_f^{3}\mathcal{C}^{\alpha\beta}_{\kappa\lambda}\delta^{\mathrm{K}}_{34} }{V^{\mathrm{t}}_{123}V^{\mathrm{t}}_{456}}
\int_{V_{\mathbf{q}_1}}\int_{V_{\mathbf{q}_2}}\int_{V_{\mathbf{q}_5}}\int_{V_{\mathbf{q}_6}}
\mathcal{M}^{(\alpha\beta\kappa\lambda)}_{\mu\nu,ab}
d^3\mathbf{q}_1d^3\mathbf{q}_2d^3\mathbf{q}_5d^3\mathbf{q}_6 \,
\delta_D\left(\mathbf{q}_1+\mathbf{q}_2-\mathbf{q}_5-\mathbf{q}_6\right)
\notag\\
&\times
\mathrm{B}_{\mathrm{g}}^{\mathrm{s}}\left(\mathbf{q}_1,\mathbf{q}_2,-\mathbf{q}_1-\mathbf{q}_2\right)\mathrm{B}_{\mathrm{g}}^{\mathrm{s}}\left(-\mathbf{q}_5-\mathbf{q}_6,\mathbf{q}_5,\mathbf{q}_6\right) \; + 8 \;\mathrm{perm.}
\notag \\
&\approx
\dfrac{k_f^{3}\mathcal{C}^{\alpha\beta}_{\kappa\lambda}\delta^{\mathrm{K}}_{34} }{V^{\mathrm{t}}_{123}V^{\mathrm{t}}_{456}}\times \dfrac{V^{\mathrm{q}}_{1256}}{8}
\int_{-1}^{+1}d\mu_3d\mu_1d\mu_5 
\mathcal{M}^{(\alpha\beta\kappa\lambda)}_{\mu\nu,ab}
\mathrm{B}_{\mathrm{g}}^{\mathrm{s}}\left(k_1,k_2,k_3,\mu_3,\mu_1\right)\mathrm{B}_{\mathrm{g}}^{\mathrm{s}}\left(k_3,k_5,k_6,\mu_3,\mu_5\right) \; + 8 \;\mathrm{perm.}
\notag \\
&=
\dfrac{k_f^{3}\mathcal{C}^{\alpha\beta}_{\kappa\lambda}\delta^{\mathrm{K}}_{34} }{4\pi k_3k_4\Delta k}\times \dfrac{1}{8}
\int_{-1}^{+1}d\mu_3d\mu_1d\mu_5 
\mathcal{M}^{(\alpha\beta\kappa\lambda)}_{\mu\nu,ab}
\mathrm{B}_{\mathrm{g}}^{\mathrm{s}}\left(k_1,k_2,k_3,\mu_3,\mu_1\right)\mathrm{B}_{\mathrm{g}}^{\mathrm{s}}\left(k_3,k_5,k_6,\mu_3,\mu_5\right) \; + 8 \;\mathrm{perm.}
,
\end{align}
\noindent where the usual approximations have been used together with Equation \eqref{eq:tri_int_vol}; in the last step these approximations have  enabled us to simplify the quadrilateral and triangles volumes ratio. Notice that in this case the $k$-vectors $\mathbf{k}_3$ and $\mathbf{k}_4$ coincide, given the implicit $\delta_D(\mathbf{q}_3-\mathbf{q}_4)$.

The non-Gaussian term proportional to the trispectrum monopole ($\propto PT$) presented below is original of this work.

\begin{align}
\label{eq:cov_pt_ng}
&\mathrm{C}_{\propto\mathrm{TP}}^{\mathrm{B}^{\alpha\beta}_{\mathrm{g}}\mathrm{B}^{\kappa\lambda}_{\mathrm{g}}}\left(k_1,k_2,k_3;k_a,k_b,k_c\right) = 
\notag \\
&=
\dfrac{(2\pi k_f)^{-6}\mathcal{C}^{\alpha\beta}_{\kappa\lambda}}{N^{\mathrm{t}}_{123}N^{\mathrm{t}}_{abc}}
\prod^{3,c}_{i=1,a}\int_{V_{\mathbf{q}_i}} d^3\mathbf{q}_i
\mathcal{M}^{(\alpha\beta\kappa\lambda)}_{\mu\nu,ab}
\delta_D\left(\mathbf{q}_1+\mathbf{q}_2+\mathbf{q}_3\right)\delta_D\left(\mathbf{q}_a+\mathbf{q}_b+\mathbf{q}_c\right)
\notag\\
&\times
(2\pi)^6\delta_D\left(\mathbf{q}_1+\mathbf{q}_2+\mathbf{q}_a+\mathbf{q}_b\right)\delta_D\left(\mathbf{q}_3+\mathbf{q}_c\right)
\mathrm{T}_{\mathrm{g}}^{\mathrm{s}}\left(\mathbf{q}_1,\mathbf{q}_2,\mathbf{q}_a,\mathbf{q}_b\right)\mathrm{P}_{\mathrm{g}}^{\mathrm{s}}\left(\mathbf{q}_3\right) \; + 8 \;\mathrm{perm.}
\notag \\
&=
\dfrac{\delta^{\mathrm{K}}_{3c}k_f^{-6}\mathcal{C}^{\alpha\beta}_{\kappa\lambda}}{N^{\mathrm{t}}_{123}N^{\mathrm{t}}_{abc}}
\int d^3\mathbf{q}_1d^3\mathbf{q}_2d^3\mathbf{q}_3d^3\mathbf{q}_ad^3\mathbf{q}_b
\mathcal{M}^{(\alpha\beta\kappa\lambda)}_{\mu\nu,ab}
\delta_D\left(\mathbf{q}_1+\mathbf{q}_2+\mathbf{q}_a+\mathbf{q}_b\right)\delta_D\left(\mathbf{q}_1+\mathbf{q}_2+\mathbf{q}_3\right)
\notag\\
&\times
\delta_D\left(\mathbf{q}_a+\mathbf{q}_b-\mathbf{q}_3\right)
\mathrm{T}_{\mathrm{g}}^{\mathrm{s}}\left(\mathbf{q}_1,\mathbf{q}_2,\mathbf{q}_a,\mathbf{q}_b\right)\mathrm{P}_{\mathrm{g}}^{\mathrm{s}}\left(\mathbf{q}_3\right) \; + 8 \;\mathrm{perm.}
\notag \\
&=
\dfrac{\delta^{\mathrm{K}}_{3c}\delta^{\mathrm{K}}_{|1+2|3}k_f^{-6}\mathcal{C}^{\alpha\beta}_{\kappa\lambda}}{N^{\mathrm{t}}_{123}N^{\mathrm{t}}_{abc}}
\int d^3\mathbf{q}_1d^3\mathbf{q}_2d^3\mathbf{q}_ad^3\mathbf{q}_b
\mathcal{M}^{(\alpha\beta\kappa\lambda)}_{\mu\nu,ab}
\delta_D\left(\mathbf{q}_1+\mathbf{q}_2+\mathbf{q}_a+\mathbf{q}_b\right)^2
\notag \\
&\times
\mathrm{T}_{\mathrm{g}}^{\mathrm{s}}\left(\mathbf{q}_1,\mathbf{q}_2,\mathbf{q}_a,\mathbf{q}_b\right)\mathrm{P}_{\mathrm{g}}^{\mathrm{s}}\left(-\mathbf{q}_1-\mathbf{q}_2\right) \; + 8 \;\mathrm{perm.}
\notag \\
&=
\dfrac{\delta^{\mathrm{K}}_{3c}k_f^{-9}\mathcal{C}^{\alpha\beta}_{\kappa\lambda}}{N^{\mathrm{t}}_{123}N^{\mathrm{t}}_{abc}}
\int d^3\mathbf{q}_1d^3\mathbf{q}_2d^3\mathbf{q}_ad^3\mathbf{q}_b
\mathcal{M}^{(\alpha\beta\kappa\lambda)}_{\mu\nu,ab}
\delta_D\left(\mathbf{q}_1+\mathbf{q}_2+\mathbf{q}_a+\mathbf{q}_b\right)
\notag\\
&\times
\mathrm{T}_{\mathrm{g}}^{\mathrm{s}}\left(\mathbf{q}_1,\mathbf{q}_2,\mathbf{q}_a,\mathbf{q}_b\right)\mathrm{P}_{\mathrm{g}}^{\mathrm{s}}\left(-\mathbf{q}_1-\mathbf{q}_2\right) \; + 8 \;\mathrm{perm.}
\notag \\
&\approx
\dfrac{\delta^{\mathrm{K}}_{3c}k_f^{3}\mathcal{C}^{\alpha\beta}_{\kappa\lambda}}{V^{\mathrm{t}}_{123}V^{\mathrm{t}}_{abc}}
\times\dfrac{V^{\mathrm{q}}_{12ab}}{8}\int^{+1}_{-1}d\mu_Dd\mu_1d\mu_b
\mathrm{T}_{\mathrm{g}}\left(k_1,k_2,k_a,k_b,D,\mu_D,\mu_1,\mu_b\right)\mathrm{P}_{\mathrm{g}}\left(D,-\mu_D\right)
\; + 8 \;\mathrm{perm.}
\notag \\
&=
\dfrac{\delta^{\mathrm{K}}_{3c}k_f^{3}\mathcal{C}^{\alpha\beta}_{\kappa\lambda}\Delta D}{4\pi k_3k_c\Delta k_3\Delta k_c}\times\dfrac{1}{8}
\int^{+1}_{-1}d\mu_Dd\mu_1d\mu_b
\mathrm{T}_{\mathrm{g}}\left(k_1,k_2,k_a,k_b,D,\mu_D,\mu_1,\mu_b\right)\mathrm{P}_{\mathrm{g}}\left(D,-\mu_D\right)
\; + 8 \;\mathrm{perm.}
\,,
\end{align}

\noindent where $\delta^{\mathrm{K}}_{|1+2|3}$ is automatically satisfied and the quadrilateral diagonal is $D\equiv k_3,k_c$. In the last passage we used the expression for the trispectrum integration volume in Fourier space given in Equation \eqref{eq:tri_int_vol}. Once more we made the approximation that the trispectrum and power spectrum monopoles do not significantly vary inside the integration volume. We choose to integrate out the angular dependence using the cosine of the angles made by the quadrilateral diagonal (shared triangle side) and two of the quadrilateral sides with respect to the line of sight. 
Notice that in this case the $k$-vectors $\mathbf{k}_3$ and $\mathbf{k}_c$ are opposite given the implicit $\delta_D(\mathbf{q}_3+\mathbf{q}_c)$.

\subsection{Power spectrum - bispectrum cross-covariance}

Analogously to  Equation \ref{eq:bk_cov_expansion} for the bispectrum  covariance, the cross-covariance between power spectrum and bispectrum multipoles is obtained expanding the five-points correlator in Fourier space:

\begin{align}
\label{eq:pkbk_cov_expansion}
\langle\delta_\mathrm{g}(\mathbf{k}_1)\delta_\mathrm{g}(\mathbf{k}_2)\delta_\mathrm{g}(\mathbf{k}_3)\delta_\mathrm{g}(\mathbf{k}_4)\delta_\mathrm{g}(\mathbf{k}_5) \rangle 
&=
\langle\delta_\mathrm{g}(\mathbf{k}_1)\delta_\mathrm{g}(\mathbf{k}_3)\rangle_{\mathrm{c}}\langle\delta_\mathrm{g}(\mathbf{k}_2)\delta_\mathrm{g}(\mathbf{k}_4)\delta_\mathrm{g}(\mathbf{k}_5)\rangle_{\mathrm{c}} \quad + \quad 5\,\mathrm{perm.}
\notag \\
&+\langle\delta_\mathrm{g}(\mathbf{k}_1)\delta_\mathrm{g}(\mathbf{k}_2)\delta_\mathrm{g}(\mathbf{k}_3)\delta_\mathrm{g}(\mathbf{k}_4)\delta_\mathrm{g}(\mathbf{k}_5)\rangle_{\mathrm{c}}
\end{align}

\noindent The above expansion returns a  term proportional to the product between power spectrum and bispectrum ($\propto\delta^6_\mathrm{m}$) and a higher order term one proportional to the tetraspectrum ($\propto\delta^8_\mathrm{m}$).
Therefore, at lowest order, we compute  the covariance between the power spectrum and bispectrum multipoles in the following way:

\begin{align}
\label{eq:cross_cov_pkbk}
&\mathrm{C}_{\propto\mathrm{PB}}^{\mathrm{P}^{\ell}_{\mathrm{g}}\mathrm{B}^{\alpha\beta}_\mathrm{g}}\left(k_1;k_2,k_3,k_4\right) = 
\notag \\
&=\dfrac{(2\pi)^{-6} k_f^{-3}\mathcal{C}_{\alpha\beta}^{\ell}}{N^{\mathrm{p}}_{1}N^{\mathrm{t}}_{234}}
\prod^{4}_{i=1}\int_{V_{\mathbf{q}_i}} d^3\mathbf{q}_i\int_{V_{\mathbf{p}_1}} d^3\mathbf{p}_1
\mathcal{M}^{(\ell\alpha\beta)}_{\mu_a\mu_b\nu_b}
\delta_D\left(\mathbf{p}_1+\mathbf{q}_1\right)
\delta_D\left(\mathbf{q}_2+\mathbf{q}_3+\mathbf{q}_4\right)
\notag\\
&\times
2(2\pi)^6
\delta_D\left(\mathbf{q}_1+\mathbf{q}_2\right)\delta_D\left(\mathbf{p}_1+\mathbf{q}_3+\mathbf{q}_4\right)
\mathrm{P}_{\mathrm{g}}^{\mathrm{s}}\left(\mathbf{q}_1\right)
\mathrm{B}_{\mathrm{g}}^{\mathrm{s}}\left(\mathbf{p}_1,\mathbf{q}_3,\mathbf{q}_4\right) \; + 2 \;\mathrm{perm.}
\notag \\
&=\dfrac{2 k_f^{-3}\mathcal{C}_{\alpha\beta}^{\ell}\delta^{\mathrm{K}}_{12}}{N^{\mathrm{p}}_{1}N^{\mathrm{t}}_{234}}
\prod^{4}_{i=2}\int_{V_{\mathbf{q}_i}} d^3\mathbf{q}_i
\mathcal{M}^{(\ell\alpha\beta)}_{\mu_a\mu_b\nu_b}
\delta_D\left(\mathbf{q}_2+\mathbf{q}_3+\mathbf{q}_4\right)^2
\mathrm{P}_{\mathrm{g}}^{\mathrm{s}}\left(\mathbf{q}_2\right)
\mathrm{B}_{\mathrm{g}}^{\mathrm{s}}\left(\mathbf{q}_2,\mathbf{q}_3,\mathbf{q}_4\right) \; + 2 \;\mathrm{perm.}
\notag \\
&=\dfrac{2 k_f^{-6}\mathcal{C}_{\alpha\beta}^{\ell}\delta^{\mathrm{K}}_{12}}{N^{\mathrm{p}}_{1}N^{\mathrm{t}}_{234}}
\prod^{4}_{i=2}\int_{V_{\mathbf{q}_i}} d^3\mathbf{q}_i
\mathcal{M}^{(\ell\alpha\beta)}_{\mu_a\mu_b\nu_b}
\delta_D\left(\mathbf{q}_2+\mathbf{q}_3+\mathbf{q}_4\right)
\mathrm{P}_{\mathrm{g}}^{\mathrm{s}}\left(\mathbf{q}_2\right)
\mathrm{B}_{\mathrm{g}}^{\mathrm{s}}\left(\mathbf{q}_2,\mathbf{q}_3,\mathbf{q}_4\right) \; + 2 \;\mathrm{perm.}
\notag \\
&\approx
\dfrac{2\mathcal{C}_{\alpha\beta}^{\ell}\delta^{\mathrm{K}}_{12}}{N^{\mathrm{p}}_{1}}
\dfrac{1}{4\pi}\int^{+1}_{-1}\int^{2\pi}_{0}d\phi\mathcal{M}^{(\ell\alpha\beta)}_{\mu_a\mu_b\nu_b}
\mathrm{P}_{\mathrm{g}}^{\mathrm{s}}\left(k_2,\mu_2\right)
\mathrm{B}_{\mathrm{g}}^{\mathrm{s}}\left(k_2,k_3,k_4,\mu_2,\phi\right) \; + 2 \;\mathrm{perm.}
\notag\\
&=
\dfrac{2\,k_f^3\mathcal{C}_{\alpha\beta}^{\ell}\delta^{\mathrm{K}}_{12}}{4\pi k_2^2 \Delta k}
\dfrac{1}{4\pi}\int^{+1}_{-1}\int^{2\pi}_{0}d\phi\mathcal{M}^{(\ell\alpha\beta)}_{\mu_a\mu_b\nu_b}
\mathrm{P}_{\mathrm{g}}^{\mathrm{s}}\left(k_2,\mu_2\right)
\mathrm{B}_{\mathrm{g}}^{\mathrm{s}}\left(k_2,k_3,k_4,\mu_2,\phi\right) \; + 2 \;\mathrm{perm.}
\end{align}{}

\noindent Notice that the $\Delta k$ appearing at the denominator comes from the number of modes available in Fourier space for the power spectrum, which can be defined differently from the one relative to the bispectrum integration volume. For the power, spectrum the multipole expansion is done in terms of $\mu_a$, the cosine of the angle between the $k$-vector and the line of sight. The angle $\phi$ is the same as the one defined below Equation \eqref{eq:cov_bb_g}.

\section{Integration volumes and number of modes}
\label{sec:appendixA}
In this Appendix we expand the work presented in Appendix A of \cite{Gualdi:2018pyw}.

\begin{figure}[tbp]
\centering 
\includegraphics[width=.98\textwidth]
{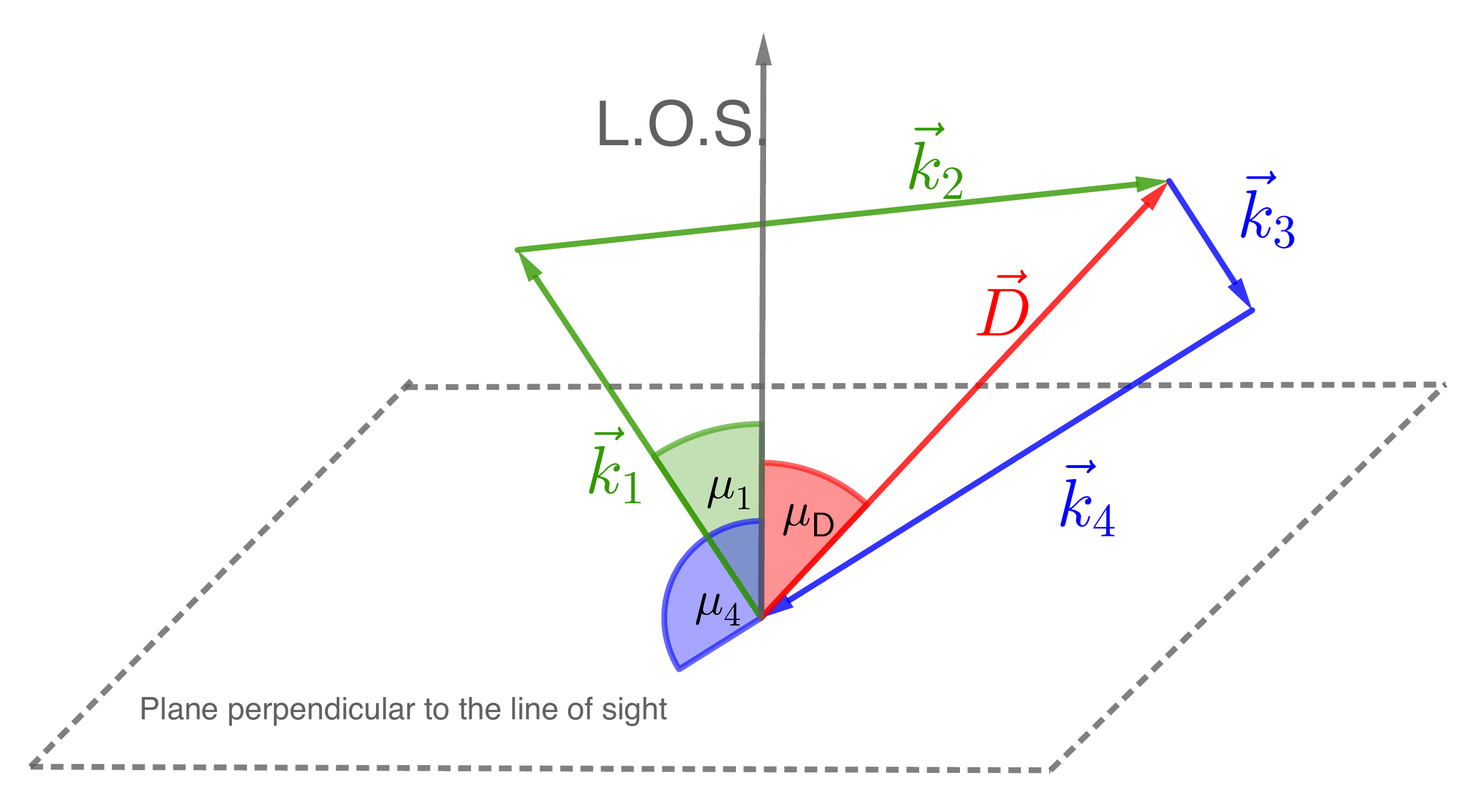}
\caption{\label{fig:coordinates_tk} trispectrum coordinates used to describe each configuration in the analytical model. $\mu_D,\mu_1,\mu_4$ are the cosine of the angles between the line of sight and the vectors $\mathbf{D},\mathbf{k}_1,\mathbf{k}_4$, respectively.}
\end{figure}

\subsection{Trispectrum integration volume and grid points quadruplets}
\label{sec:trispectrum}
The key to compute the integration volume for the trispectrum consists in considering one of the two diagonals. By doing so each trispectrum configuration can be described by 8 variables (12 degrees of freedom minus 3 for the Dirac's delta closure condition minus 1 for the invariance with respect rotations around the line of sight). A good choice for these are the four sides modules $\left(k_1,k_2,k_3,k_4\right)$, one of the two diagonals $D$ (let's choose the one defining the triangles $\overline{k_1k_2D}$ and $\overline{k_3k_4D}$ with closure conditions $\delta_D\left(\mathbf{k}_1+\mathbf{k}_2-\mathbf{D}\right)$ and $\delta_D\left(\mathbf{k}_3+\mathbf{k}_4+\mathbf{D}\right)$ respectively) and the cosine of three angles with respect to the line of sight $\left(\mu_D,\mu_1,\mu_4\right)$ for the diagonal $\mathbf{D}$ and the two sides $\mathbf{k}_1$, $\mathbf{k}_4$.  For a visual representation of the choice of coordinates see Figure \ref{fig:coordinates_tk}.

All the other details can be obtained using as coordinates the parallel/perpendicular components of the k-vectors with respect to  the line of sight. In this way, to fully define the vector, we must specify also the azimuthal angle on the plane perpendicular to the line of sight. This allows  us to fully exploit the invariance symmetry related to the rotation of the quadrilateral around the line of sight. This is done by setting the origin of the azimuthal angle  to be in the diagonal  ($\gamma_D=0$) and by computing the other four azimuthal angles with respect to  it, using the condition that the projection on the perpendicular plane must also be a closed quadrilateral composed by two closed triangles.

Considering the two triangles mentioned above, we start from  the diagonal vector $\mathbf{D}$ yielding the initial volume $V_D\simeq 4\pi D^2\Delta D$.
Now we can proceed for each triangle as in the case of the bispectrum integration volume (see Appendix of \cite{Gualdi:2018pyw}) obtaining the symmetric final result:

\begin{eqnarray}
\label{eq:tri_int_vol}
V_{\mathrm{q}} = V_D \times V_{k_{12}} \times V_{k_{34}} =16\pi^3 k_1k_2k_3k_4\Delta k_1\Delta k_2\Delta k_3\Delta k_4 \Delta D \,.
\end{eqnarray}{}

\noindent The number of grid points quadruplets is then defined as $N_{\mathrm{q}}=\dfrac{V_{\mathrm{q}}}{k_f^9}$.

\subsection{Tetraspectrum integration volume and grid points quintuplets}

For future reference we also include the integration volume for the tetraspectrum. This can be derived using the same trick of splitting the figure into triangles made by the sides and the diagonals. A particular pentagon configuration is described by 11 variables (15 degrees of freedom minus 3 for the Dirac's delta closure condition minus 1 for the invariance with respect rotations around the line of sight). 

Once more we choose the 5 sides modules $\left(k_1,k_2,k_3,k_4,k_5\right)$ and two diagonals $D_A$ and $D_B$ such that the resulting triangles are $\overline{D_Ak_2k_3}$, $\overline{D_Ak_1D_B}$ and $\overline{D_Bk_4k_5}$. 
Finally we use the four cosine of the angles with respect to the line of sight $\left(\mu_{D_A},\mu_{D_B},\mu_2,\mu_4\right)$.

The volume can be computed starting from the one for the quadrilateral $\overline{D_Bk_1k_2k_3}$ which is:

\begin{eqnarray}
V^{B123}_{\mathrm{q}} = 16\pi^3D_Bk_1k_2k_3 \Delta D_B\Delta k_1\Delta k_2\Delta k_3\,.
\end{eqnarray}{}
\noindent Again applying the bispectrum formula for the triangle $\overline{D_Bk_4k_5}$ we have that:

\begin{eqnarray}
V_{k_{45}} = 2\pi \dfrac{k_4k_5}{D_B}\Delta k_4\Delta k_5\,,
\end{eqnarray}{}
\noindent and therefore the tetraspectrum integration volume $V_{\mathrm{c}}$ ('c' stands for cinco/cinque) can be obtained by simply multiplying

\begin{eqnarray}
\label{eq:tet_int_vol}
V_{\mathrm{c}} = V^{B123}_{\mathrm{q}}\times V_{k_{45}} = 32\pi^4k_1k_2k_3k_4k_5\Delta k_1\Delta k_2\Delta k_3\Delta k_4\Delta k_5\Delta D_A \Delta D_B\,.
\end{eqnarray}

\noindent The number of grid points quintuplets is then defined as $N_{\mathrm{c}}=\dfrac{V_{\mathrm{c}}}{k_f^{12}}$.

\section{Survey window effect on the covariance matrix}
\label{sec:window_effect}

In this Appendix we investigate the origin of the features parallel to the main diagonal of the numerically estimated covariance matrix in Figure \ref{fig:covariance_comparison2D}. For this purpose we use the measurements of the same data-vector (power spectrum monopole and quadrupole plus bispectrum monopole) made on simulated cubic boxes with periodic boundary conditions \cite{Chuang:2014vfa}. Because of the fewer number of measurements available (440 with respect to 1400), the bispectrum data-vector contains 47 triangle configurations, less than the 68 employed in Figure \ref{fig:covariance_comparison2D}.

\begin{figure}[tbp]
\centering 
\includegraphics[width=.98\textwidth]
{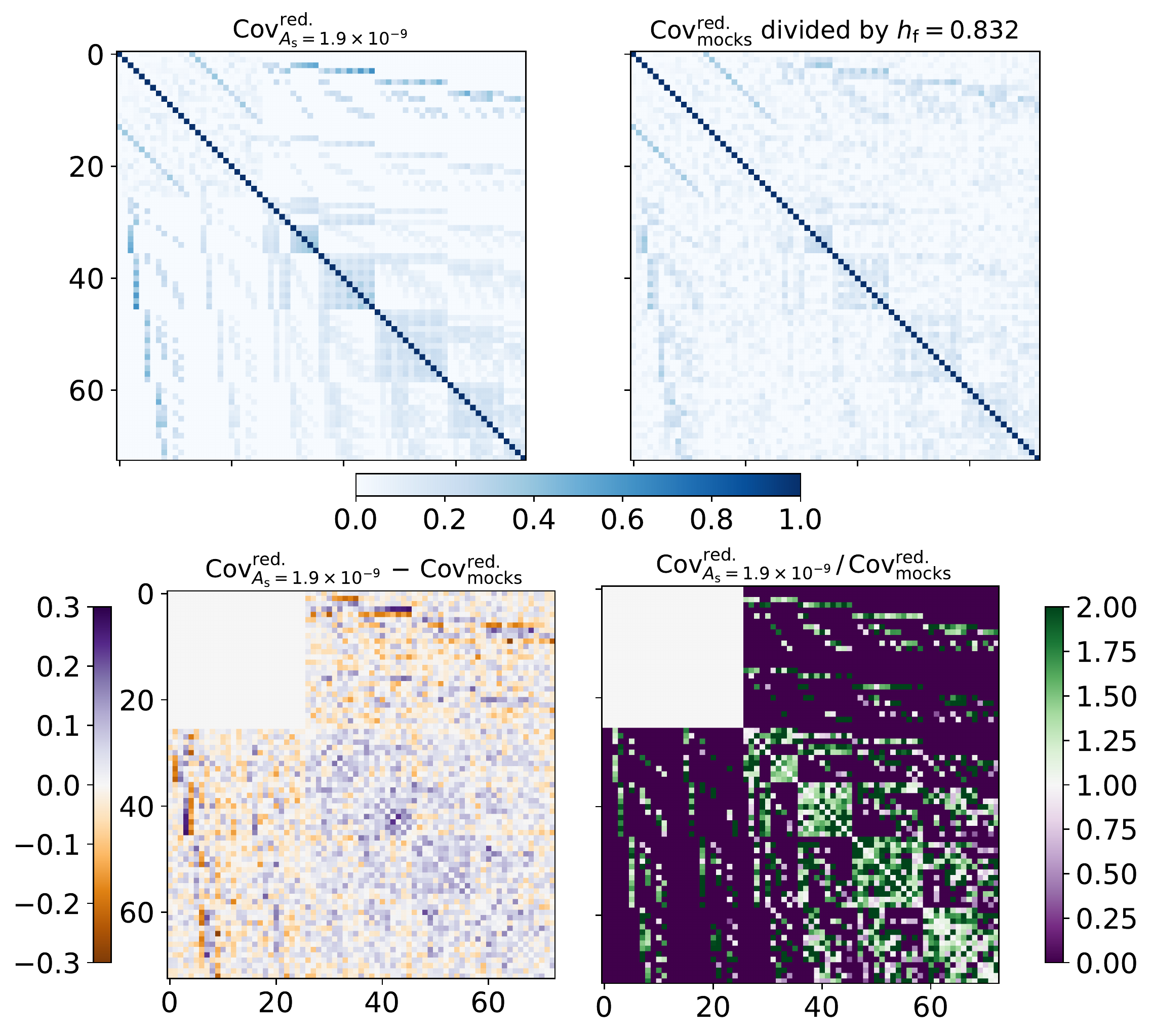}
\caption{\label{fig:covariance_comparison2D periodic}  Same as Figure \ref{fig:covariance_comparison2D} but with the numerical covariance obtained by measuring the power spectrum (monopole and quadrupole) and bispectrum monopole from a set of 440 simulated cubic volumes with periodic boundary conditions. In Figure \ref{fig:covariance_comparison2D} we used 1400 measurements from galaxy mock catalogues to estimate the covariance matrix for the bispectrum of 68 triangles configurations. Here we considered 47 triangle configurations to derive the bispectrum covariance  from 440 simulated cubic boxes with periodic boundary conditions. The overall discrepancy between the analytical and numerical covariances described in the two bottom panels is qualitatively the same as the one in Figure \ref{fig:covariance_comparison2D}. However in this case where the galaxy field has a regular geometry (cubic box with side $L=5000\, h/$Mpc ) and therefore no survey window selection effect, those features parallel to the main diagonal present in Figure \ref{fig:covariance_comparison2D} are here absent.
Therefore this test corroborates our hypothesis of those features being related to the mode-coupling induced by the survey window selection effect.
  }
\end{figure}

While the overall discrepancy between analytical and numerical estimates is approximately the same as in Figure \ref{fig:covariance_comparison2D}, in the top-right panel of Figure \ref{fig:covariance_comparison2D periodic} there are no features parallel to the main diagonal. This difference between Figure \ref{fig:covariance_comparison2D} and Figure \ref{fig:covariance_comparison2D periodic} supports our hypothesis that those features are not physical but related to the mode-coupling induced by the survey window selection function.

\section{\texorpdfstring{$\mathbf{\Delta k_6}$}{} binning case figures}
\label{sec:deltak6}
This last Appendix contains the two key figures and improvements table for the $\Delta k_6$ binning case.

\begin{figure}[tbp]
\centering 
\includegraphics[width=.98\textwidth]
{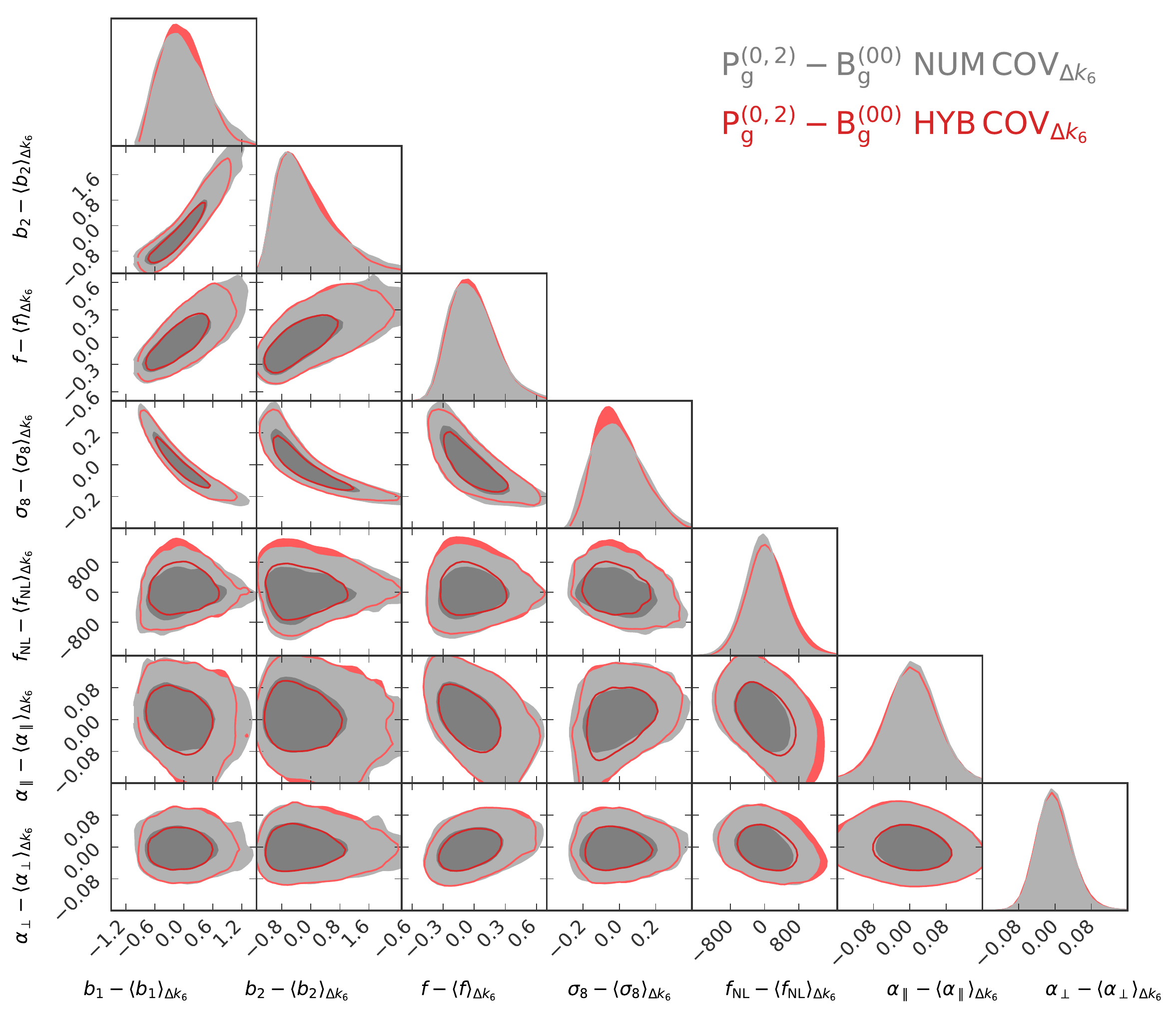}
\caption{\label{fig:p02b00p6dk6}
Same as Figure \ref{fig:p02b00p6dk5} for the $\Delta k_6$ binning case.}
\end{figure}

\begin{figure}[tbp]
\centering 
\includegraphics[width=.98\textwidth]
{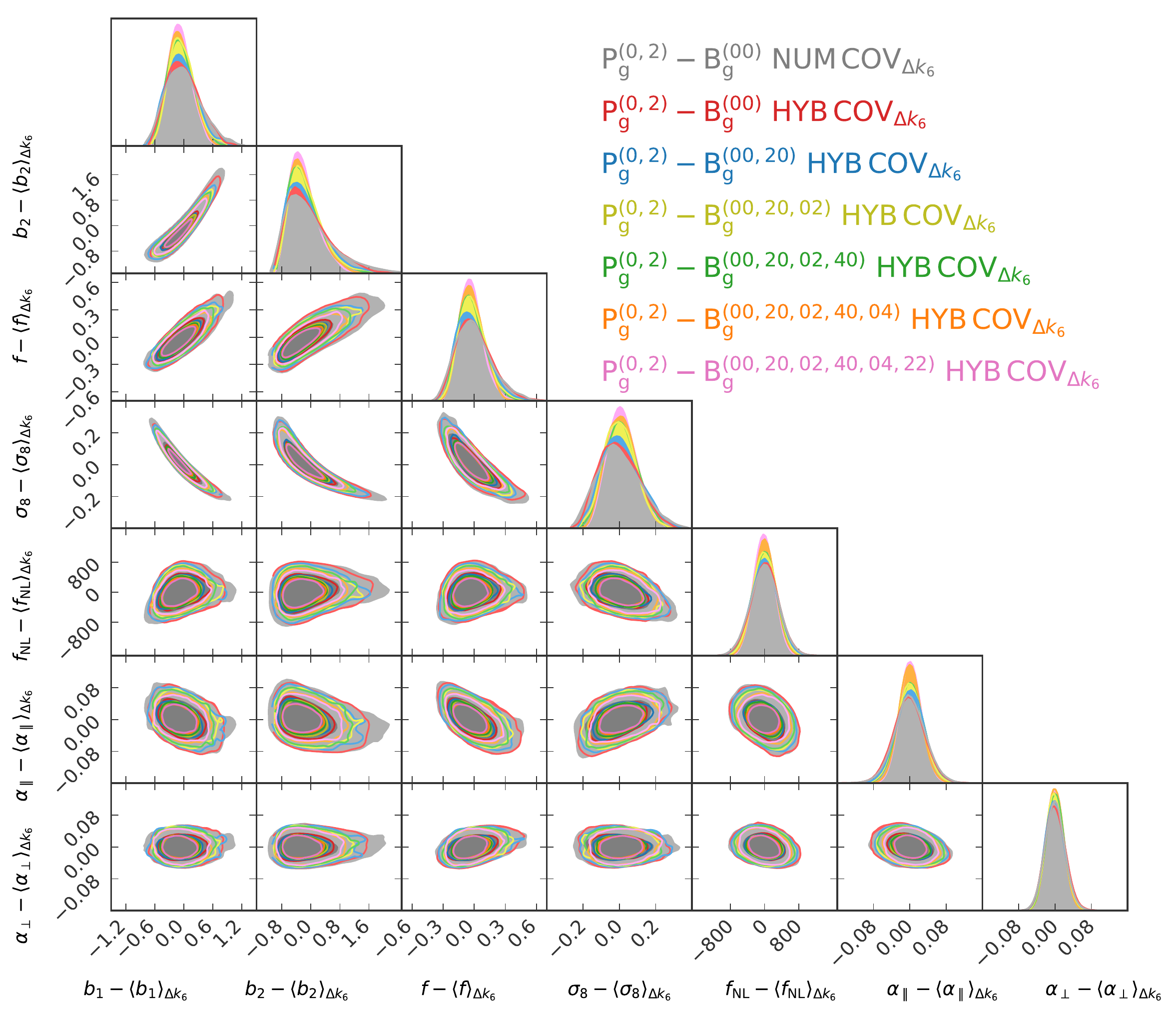}
\caption{\label{fig:ana_dv_6p_d6_improv02} Same as Figure \ref{fig:ana_dv_6p_d5_improv02} for the $\Delta k_6$ binning case.}
\end{figure}

\renewcommand{\arraystretch}{2}
\begin{table}[tbp]
\centering
\begin{tabular}{|c|cccccccc|}
\hline
statistics &  \multicolumn{8}{c|}{$\dfrac{ \Delta\theta^{\mathrm{num.\,cov.}}_{\mathrm{P_g^{(0,2)};B_g^{(00)}}}-\Delta\theta^{\mathrm{hyb.\,cov.}}_{\mathrm{P_g^{(0,2)};B_g^{(\imath,\jmath)}}}}{\Delta\theta^{\mathrm{num.\,cov.}}_{\mathrm{P_g^{(0,2)};B_g^{(00)}}}}\;\left[\%\right]$} \\ [0.5cm]
\hline
$\mathrm{P}_\mathrm{g}^{(0,2)}\;+$  & $b_1$ & $b_2$ & $f$  
& $\sigma_8$ & $f_\mathrm{NL}$  & $\alpha_\parallel$ & $\alpha_\perp$ & average\\ 
\hline   
$\mathrm{\mathrm{B}_\mathrm{g}^{(00)}}$
& 2.4 &  3.4 & -4.9 &  2.9 &  2.0 &  0.6 & -5.5 &  0.1	 \\ 
$\mathrm{\mathrm{B}_\mathrm{g}^{(00,20)}}$
& 46.8 & 53.6 & 39.4 & 31.1 & 37.6 & 43.0 & 42.8 & 42.0\\ 
$\mathrm{\mathrm{B}_\mathrm{g}^{(00,20,02)}}$
&55.8 & 61.5 & 49.6 & 43.0 & 44.0 & 50.1 & 45.0 & 49.9	\\ 
$\mathrm{\mathrm{B}_\mathrm{g}^{(00,20,02,40)}}$
&55.3 & 61.2 & 51.1 & 41.5 & 44.2 & 52.7 & 46.7 & 50.4\\ 
$\mathrm{\mathrm{B}_\mathrm{g}^{(00,20,02,40,04)}}$
& \textbf{57.8} & \textbf{63.1} & 56.9 & \textbf{45.4} & 49.8 & 59.6 & 50.8 & 54.8 \\ 
$\mathrm{\mathrm{B}_\mathrm{g}^{(00,20,02,40,04,22)}}$
& 57.1 & 62.8 & \textbf{57.0}  & 44.0 & \textbf{51.9} & \textbf{60.2} & \textbf{52.6} & \textbf{55.1}  \\ 
\hline
\end{tabular}
\caption{Same as Table \ref{tab:ana_dv_6p_d6_improv} for the $\Delta k_6$ binning case.  The table tells us that even if the average improvement is larger with respect to the $\Delta k_5$ binning case, the information gain reaches saturation always at the bispectrum hexadecapole level. At the same time if we compare the 1-2D marginalised posterior distributions shown in Figures \ref{fig:ana_dv_6p_d5_improv02} and \ref{fig:ana_dv_6p_d6_improv02} for the $\Delta k _5$ and $\Delta k _6$ cases respectively, we can see that once all the hexadecapole terms have been added to the data vector, the width of the contours is similar. 
}
\end{table}

\newpage

\acknowledgments

D.G. thanks H\'ector Gil Mar\'in for providing the data vector's measurements from the galaxy mocks and for useful discussions. D.G. and L.V. are very grateful to Cheng Zhao and Chia-Hsun Chuang for the periodic simulations used to test the survey window effect on the covariance. L.V.  and D.G. acknowledge support of European Unions Horizon 2020 research and innovation programme ERC (BePreSySe, grant agreement 725327). Funding for this work was partially provided by the Spanish Ministerio de Ciencia y Innovation y Universidades under project PGC2018-098866-B-I00.



\bibliographystyle{mnras}

\bibliography{bkquad_ref.bib}
 






\end{document}